\documentclass{svjour3}                     
\usepackage{epsf}

\begin{document}

\title{Analysis of a network structure of the foreign currency exchange 
market}

\author{Jaros{\l}aw Kwapie\'n \and Sylwia Gworek \and Stanis{\l}aw 
Dro\.zd\.z \and Andrzej G\'orski}

\institute{S. Dro\.zd\.z \at Institute of Nuclear Physics, Polish Academy 
of Sciences, ul. Radzikowskiego 152, 31-342 Krak\'ow, Poland
\at Institute of Physics, University of Rzesz\'ow, ul. Rejtana 16A,  
35-959 Rzesz\'ow, Poland
\and
Andrzej G\'orski \at Institute of Nuclear Physics, Polish Academy  
of Sciences, ul. Radzikowskiego 152, 31-342 Krak\'ow, Poland
\and
Sylwia Gworek \at Institute of Nuclear Physics, Polish Academy of 
Sciences, ul. Radzikowskiego 152, 31-342 Krak\'ow, Poland
\and
Jaros{\l}aw Kwapie\'n \at Institute of Nuclear Physics, Polish Academy  
of Sciences, ul. Radzikowskiego 152, 31-342 Krak\'ow, Poland
}

\date{Received: date / Accepted: date}

\maketitle

\begin{abstract}

We analyze structure of the world foreign currency exchange (FX)  market
viewed as a network of interacting currencies. We analyze daily time
series of FX data for a set of 63 currencies, including gold, silver and
platinum. We group together all the exchange rates with a common base
currency and study each group separately. By applying the methods of
filtered correlation matrix we identify clusters of closely related
currencies. The clusters are formed typically according to the economical
and geographical factors. We also study topology of weighted minimal
spanning trees for different network representations (i.e., for different
base currencies) and find that in a majority of representations the
network has a hierarchical scale-free structure. In addition, we analyze
the temporal evolution of the network and detect that its structure is
not stable over time. A medium-term trend can be identified which affects 
the USD node by decreasing its centrality. Our analysis shows also an 
increasing role of euro in the world's currency market.


\end{abstract}

\section{Introduction}

There are at least two reasons for analyzing the global foreign exchange 
(FX) market. First, this is the world's largest and most important 
financial market, completely decentralized, extending over all the 
countries, with the highest daily trading volume reaching trillions of US 
dollars. Second, the FX market's dynamics seems to be more complex than 
any other market's. The absence of an independent reference frame makes 
the absolute currency pricing difficult or even impossible: one has to 
express a given currency's value by means of some other currency which, in 
turn, is also denominated only in currencies. Moreover, apart from its 
internal dynamics, the global nature of the FX market implies sensitivity 
to current situation on other markets in all parts of the world. These 
properties together with the triangle rule~\cite{aiba02} which links 
mutual exchange rates of three currencies are among the factors 
responsible for a highly correlated structure of Forex.

Correlations allow one to view the FX market's structure as a network of
interacting exchange rates. In this case the exchange rates are treated as
network nodes and are linked with their neighbours via edges with weights
proportional to the coupling strength. And although the exact nature of
these interactions remains unexplained, it is justified to assume that
they are strongly nonlinear. This indicates that the FX market may
actually constitute a complex network.

An analysis of the currency exchange network can provide us with knowledge
of the structure of the market and about a role played in it by each
particular currency. We put stress on quantification of a currency's
importance in the world financial system and on tracking its subtle
changes as the market evolves in time. We achieve this by employing the
well-known methods of correlation matrix (CM) and minimal spanning trees
(MST). However, one has to be aware that both these methods, although 
simple and effective, are linear and thus they detect only a part of 
interactions between the exchange rates; nonlinear contributions to the 
internode couplings are neglected.

An exchange rate assigns value to a currency X by expressing it in terms
of a base currency B. In general, each currency can be a base for all 
other ones. Since different currencies may have different internal 
dynamics related to domestic economy, inflation, and sensitivity to events 
in other countries and markets, behaviour of the exchange rates is 
strongly dependent on a particular choice of the base. What follows, there 
is no absolute correlation structure of the FX network; its structure 
depends largely on the base currency.

\section{Methodology and Results}

\subsection{Data and nomenclature}

We analyze daily data~\cite{sauder} for a 63-element set comprising 60 
actual currencies and 3 precious metals: gold, silver, and platinum. We 
consider the inclusion of these metals in our analysis as justified due to 
the two following reasons: First, gold and other precious metals are 
historically closely related to the currency system (silver and gold 
coins, the gold standard etc.). Even if at present there is no explicit 
relation between the official monetary system and the precious metals, 
they are still perceived by many as a convenient alternative to real 
currencies in times of high inflation or deep crises. Second, we prefer to 
include the precious metals also because they, if treated as a reference 
frame, can allow us to look at the actual currency market from outside. In 
this context the precious metals can serve as a benchmark of being 
decoupled from the market.

For denoting the currencies we adopted the ISO 4217 standard using 
three-letter codes (CHF, GBP, USD etc.). Our data spans the time period of 
9.5 years from 1 January 1999 to 30 June 2008. At a time instant $t$, the 
exchange rate B/X is $R_{\rm X}^{\rm B}(t)$. We conventionally define the 
exchange rate returns $G_{\rm X}^{\rm B}(t,\Delta t)$ as the logarithmic 
exchange rate increments over an interval $\Delta t = 1$ day. From our 
basket of $N = 63$ currencies we obtained $N(N-1) = 3906$ time series 
${\bf G}_{\rm X}^{\rm B} = \{G_{\rm A}^{\rm B}(t_i)\}_{i=1,...,T}$ of 
length $T = 2394$. All time series were preprocessed in order to eliminate 
artifacts and too extreme data points that can misleadingly dominate the 
outcomes; no points that deviate more than 10 standard deviations from the 
mean were allowed.  Owing to a high quality of data, only a few data 
points in total were modified accordingly.

Dealing with all the available exchange rates for 63 currencies
simultaneously would be rather inefficient due to information overload and
would lead to results whose interpreting might be cumbersome. Attempts in
this direction can be found elsewhere - for example, in
ref.~\cite{mcdonald05}; we prefer here a more selective approach. An
indirect but useful way to get some insight into properties of an
individual currency, if only a set of its exchange rates is available, is
to single out those rates in which this currency serves as a base currency
and apply a statistical approach to the data. By selecting the base
currency one associates a reference frame with this currency. Thus, the
evolution of all other currencies expressed by relevant exchange rates is
the evolution in the frame in which the base currency ``rests''. In this
context, a statistical analysis of the exchange rates offers information
on how the global FX market looks like from the perspective of the base
currency or, conversely, how the base currency behaves in relation to the
global market.

\subsection{Correlation matrix formalism}

Details of the basic correlation matrix formalism are as follows. For a 
set of exchange rates sharing the same base B we calculate an $N \times N$ 
correlation matrix ${\bf C}^{\rm B}$:
\begin{equation}
{\bf C}^{\rm B} = {1 \over T} {\bf M}^{\rm B} \bar{ {\bf M}}^{\rm B},
\end{equation}
where ${\bf M}$ is $N \times T$ data matrix and the bar denotes matrix 
transpose. Each entry $C_{\rm X,Y}^{\rm B}$ is the correlation coefficient 
calculated for a pair of the exchange rates B/X and B/Y. If the exchange 
rates are considered as network nodes, the correlation matrix is 
equivalent to the weights matrix collecting the weights of links between 
the nodes. The so-defined correlation matrix is a starting point for our 
further calculations.

The market global correlation structure, as it is viewed from B, can be 
described by the eigenspectrum of ${\bf C}^{\rm B}$. Complete set of the 
corresponding eigenvectors ${\bf v}_i^{\rm B}$ and eigenvalues 
$\lambda_i^{\rm B}$ ($i=1,...,N-1$) can be obtained by solving the 
equation
\begin{equation}
{\bf C}^{\rm B} {\bf v}_i^{\rm B} = \lambda_i^{\rm B} {\bf v}_i^{\rm B}.
\end{equation}

For a stock market it is typical that the associated correlation matrix 
can be decomposed into three components:
\begin{equation}
{\bf C}^{\rm B} = {\bf C}_m^{\rm B} + {\bf C}_s^{\rm B} + {\bf C}_r^{\rm 
B}.
\label{decomposition}
\end{equation}
The first component ${\bf C}_m^{\rm B}$ describes a collective (market) 
mode characterizing the average behaviour of the whole market. ${\bf 
C}_s^{\rm B}$ describes the sectored structure of the market and ${\bf 
C}_r^{\rm B}$ expresses independent behaviour of individual stocks. On a 
matrix level, ${\bf C}_m^{\rm B}$ has all its entries equal and its rank 
is 1, while ${\bf C}_r^{\rm B}$ is a random matrix drawn from the Wishart 
ensemble. The middle term in Eq.~(\ref{decomposition}) is a matrix with a 
typical rank $1 \le R \le 10$ which contains the most interesting 
information on the stock market structure~\cite{plerou02,utsugi04,kim05}.

\subsection{Eigenspectra of CM for different base currencies}

Taking outcomes of some earlier works~\cite{mcdonald05,mizuno06,%
naylor07,drozdz07a} into consideration we are justified to assume that 
also the FX market possesses similar correlation structure which can be
decomposed into the analogous three levels of currency dependencies.
To inspect this, we calculate the CMs and derived their eigenvalue spectra 
for all 63 base currencies. In each case the same scheme is reproduced: 
there is a collective mode represented by the largest eigenvalue 
$\lambda_1^{\rm B}$ with $\lambda_1^{\rm B} \gg \lambda_2^{\rm B}$. Size 
of the gap between $\lambda_1^{\rm B}$ and $\lambda_2^{\rm B}$ is 
B-dependent. A few examples can be seen in Figure 1.

\begin{figure}
\epsfxsize 13cm
\epsffile{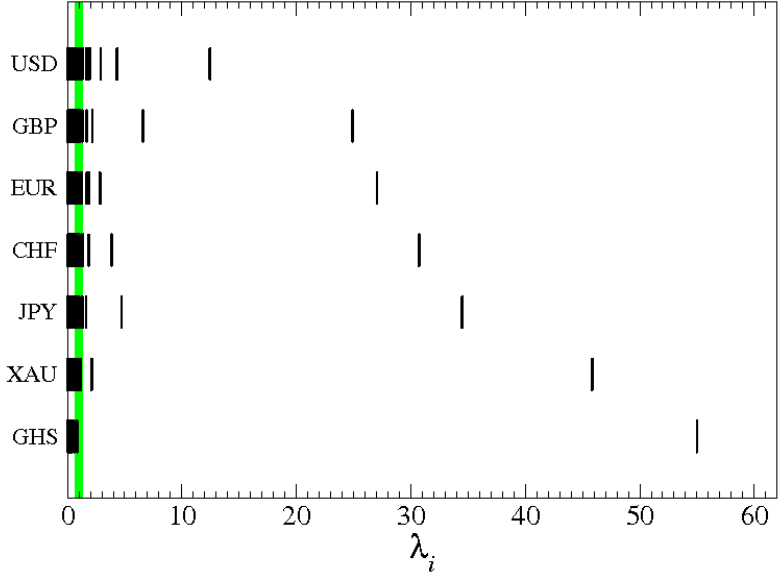}
\caption{Eigenvalue spectra of correlation matrices calculated for a few 
exemplary choices of base currency. Note different gaps between the 
largest and the second largest eigenvalues in each case. Shaded region 
corresponds to the Wishart ensemble of random matrices.}
\end{figure}

Magnitude of $\lambda_1^{\rm B}$ expresses how many exchange rates are 
correlated among themselves, i.e. how collective is the market. Properties 
of the matrix trace impose bounds on the magnitude of the largest 
eigenvalue: $1 \le \lambda_1^{\rm B} \le {\rm Tr} ({\bf C}^{\rm B}) = N - 
1$. In our case the range of the actual variability of $\lambda_1^{\rm B}$ 
is narrower: $12.29 \le \lambda_1^{\rm B} \le 55.0$ with the extrema 
reached for HKD and GHS, respectively. Figure 2 displays the corresponding 
$\lambda_1^{\rm B}$ for each B analyzed.  The whole set of currencies has 
been divided there into four baskets with respect to liquidity of each 
currency. The most important and liquid currencies belong to Basket 1 and 
other liquid ones to Basket 2.

A straightforward interpretation of $\lambda_1^{\rm B}$ points out to the 
fact that the larger it is, the more coupled is behaviour of the 
underlying exchange rate sets. That is, for large $\lambda_1^{\rm B}$ 
comparable with $N - 1$, the global FX market evolves collectively in the 
reference frame of B. This actually means that the evolution of B is 
significantly decoupled and has its own independent dynamics not related 
to the global market. In such a case the influence of this currency on 
other currencies is marginal if at all. What is natural, the precious 
metals (XAG, XAU, XPT) as commodities qualify here, but, surprisingly, the 
same is true for a few actual currencies (e.g. GHS, DZD, ZAR, BRL). 
Reasons for this type of behaviour may comprise high inflation rate in the 
corresponding countries or a strong regulation of the market by local 
financial authorities (Basket 4). It is worth noting that no Basket 1 
currency belongs to this group.

On the opposite pole (relatively small $\lambda_1^{\rm B}$) there is the
US dollar and a few other currencies from different baskets (CNY, HKD, SGD
etc.) In general, small values of $\lambda_1$ are developed by the systems
which do not display strong couplings among its elements. Hence, evolution
of the exchange rates seen from the USD perspective must be rather
decorrelated and many currencies enjoy large amounts of independence. Such
a degree of independence is not observed for any other liquid,
market-valued base currencies from Basket 1. Thus, by changing the base
from one of those currencies to USD, many explicit satellites of USD and
its more delicately related companions acquire a dose of freedom. This
phenomenon is a manifestation of the leading role of USD in the global
foreign exchange system.

A careful investigation of Figure 2 suggests that healthy, freely
convertible currencies generally are associated with $\lambda_1^{\rm B} < 
40$.

\begin{figure}
\epsfxsize 13cm
\epsffile{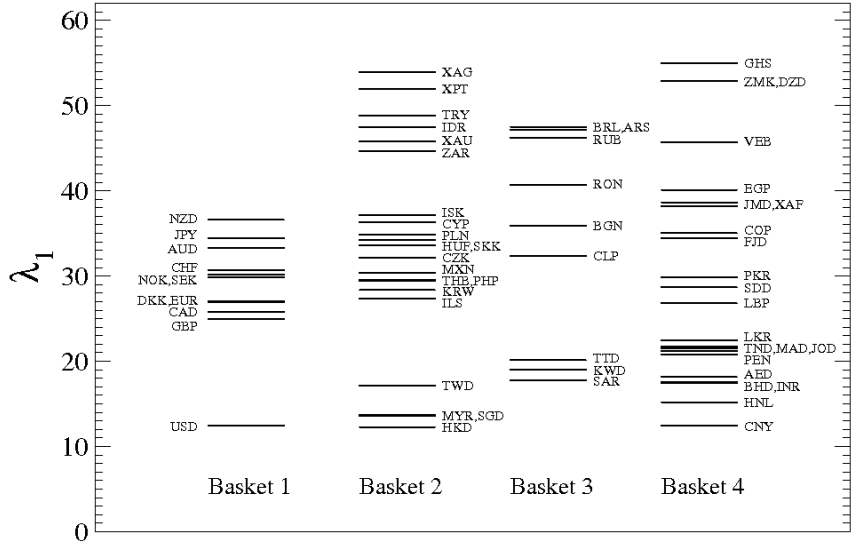}
\caption{The largest eigenvalue $\lambda_1^{\rm B}$ for different base 
currencies. All 63 currencies are distributed among 4 baskets defined by 
a currency's liquidity and its freedom of trading. Basket 1 contains the 
most liquid, freely convertible currencies, while Basket 4 contains only
fully regulated and/or non-convertible ones.}
\label{fig:2}
\end{figure}

Going back to Figure 1, it is evident from the eigenspectra that the FX 
market might have a finer sectored structure similarly to the stock and 
the commodity markets~\cite{sieczka08}. For most base currencies, apart 
from $\lambda_1^{\rm B}$ there are smaller eigenvalues which also do not 
coincide with the spectrum predicted for the Wishart matrix ensemble by 
the random matrix theory~\cite{sengupta99} (see the shaded region in 
Figure 1). A possible way to extract information on a more subtle 
structure of the FX market would be removing the market component ${\bf 
C}_m^{\rm B}$ from the matrix ${\bf C}^{\rm B}$, since it absorbs a 
significant fraction of the total variance of signals and suppresses other 
components. However, we prefer here an alternative approach, based on 
direct removing of specific exchange rates.

It is well known that due to strength of the associated economies and the
investors confidence, the US dollar and euro are the most influential
currencies. Their significant impact on other currencies is manifested in
the network representation of the FX market by a key positions of the
nodes representing the exchange rates involving at least one of these
currencies~\cite{mizuno06,naylor07,gorski08}. For all choices of B,
satellite currencies of either USD or EUR have their exchange rates
strongly correlated with B/USD or B/EUR. This implies that, from a point 
of view of a given base currency, the collective behaviour of the market 
expressed, e.g., in terms of the repelled $\lambda_1^{\rm B}$ is, at least 
in part, an effect of these correlations. In order to get more insight 
into subtle dependencies among the exchange rates, now masked by the USD- 
and EUR-induced correlations, these two groups of couplings have to be 
removed. This can be accomplished by subtracting the USD- and EUR-related 
components from each original signal B/X by least square fitting 
$G_{\rm Y}^{\rm B}(t)$ to $G_{\rm X}^{\rm B}(t)$ (Y denotes either USD or 
EUR):
\begin{equation}
G_{\rm X}^{\rm B} (t) = a G_{\rm Y}^{\rm B}(t) + b + \epsilon_{\rm X}^{\rm 
B}(t), \ \ \ i = 1,...,N - 1.
\label{removal}
\end{equation}
The residual component $\epsilon_{\rm X}^{\rm B}(t)$ is just the exchange 
rate (B/X)' linearly independent of B/Y. 

\begin{figure}
\epsfxsize 13cm
\epsffile{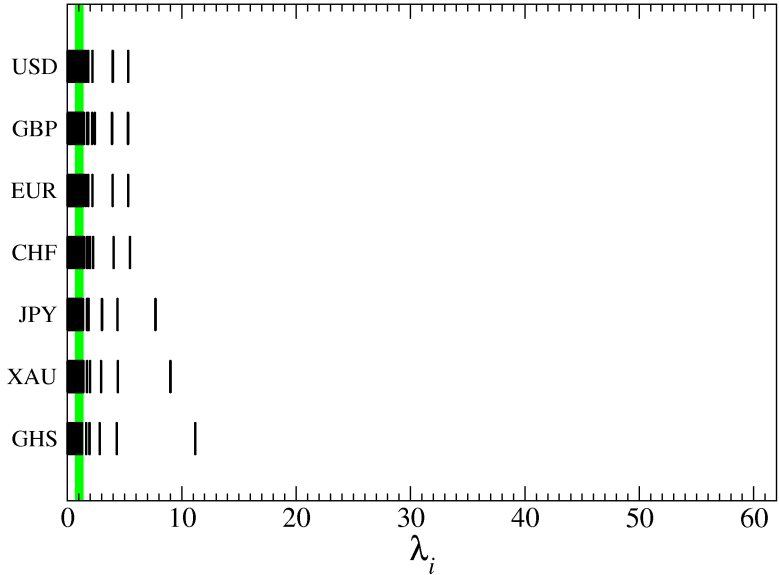}
\caption{Eigenvalue spectra of correlation matrices ${\bf C}^{\rm B}$
after removing contributions from B/USD and B/EUR cross-rates. The spectra
for the same base currencies as in Figure 1 are shown. Shaded region 
corresponds to the Wishart ensemble of random matrices.}
\label{fig:3}
\end{figure}

For a given B, we apply the above procedure to each exchange rate B/X
twice: first for Y=USD, then for Y=EUR. Of course, if B=USD or B=EUR,
Eq.(\ref{removal}) is applied only once, since the cases of the EUR/EUR
and USD/USD rates are trivial. The correlation matrix ${\bf C}^{\rm 'B}$
constructed from the signals $\epsilon_{\rm X}^{\rm B}(t)$ can again be
diagonalized and its eigenspectrum can be calculated. Figure 3 shows the
so-modified eigenspectra for the same base currencies as in Figure 1.
Clearly, now there is no collective market mode for USD, EUR, GBP and CHF.
This means that, for this group of base currencies, the entire
collectivity of the market, as seen in the corresponding values of
$\lambda_1^{\rm B}$ in Figure 2, stems from the couplings between USD and
its satellites and between EUR and its satellites. No other factor
contributes here. The eigenspectra with some eigenvalues which still do
not fit into the predicted range for the random matrices do not show any
considerable differences between different base currencies. This
observation, however, is not valid for JPY, XAU and GHS (Figure 3). Here
the largest eigenvalue deviates more from the rest of the spectrum,
although not so strongly as in Figure 1. This residual collective 
behaviour can be explained by the existence of a significantly populated 
cluster of currencies which survived the process of removing the B/USD and 
B/EUR.

\subsection{Cluster structure of the FX market}
\label{clusters}

The residual correlations which develop the non-random structure of the CM
eigenspectra in Figure 3 lead to the existence of currency clusters which
previously were masked by the dominating nodes of USD and EUR. 
Identification of these clusters can be carried out with help of a simple 
method of discriminating threshold applied to CM entries~\cite{kim05}. 
The procedure is as follows: we stepwise change the threshold $p$ 
from its maximal value $p=1$ down to 0 or even below 0. For each value of 
$p$ we preserve only those matrix entries that obey the inequality 
$C_{X,Z}^{'B} > p$ and substitute zeros otherwise. Then we count the 
clusters of at least two exchange rates. We define clusters as the 
disjoint sets of all residual exchange rates (B/X)' that are linked by 
a non-zero matrix entries to at least one cross-rate (B/Z)', Z$\ne$X.

Obviously, for $p=1$ there is no cluster and for sufficiently small $p$
there is exactly one cluster comprising all $N-1$ exchange rates. On the 
other hand, for the intermediate values of $p$ the number of clusters 
varies and may exceed 1. In order to identify the finest possible cluster 
structure of the market and to identify the exchange rates that form each 
cluster, we fix the threshold at $p=p_c$ for which the number of clusters 
is stable and close to a maximum.

\begin{figure}
\hspace{0.1cm}
\epsfxsize 6cm
\epsffile{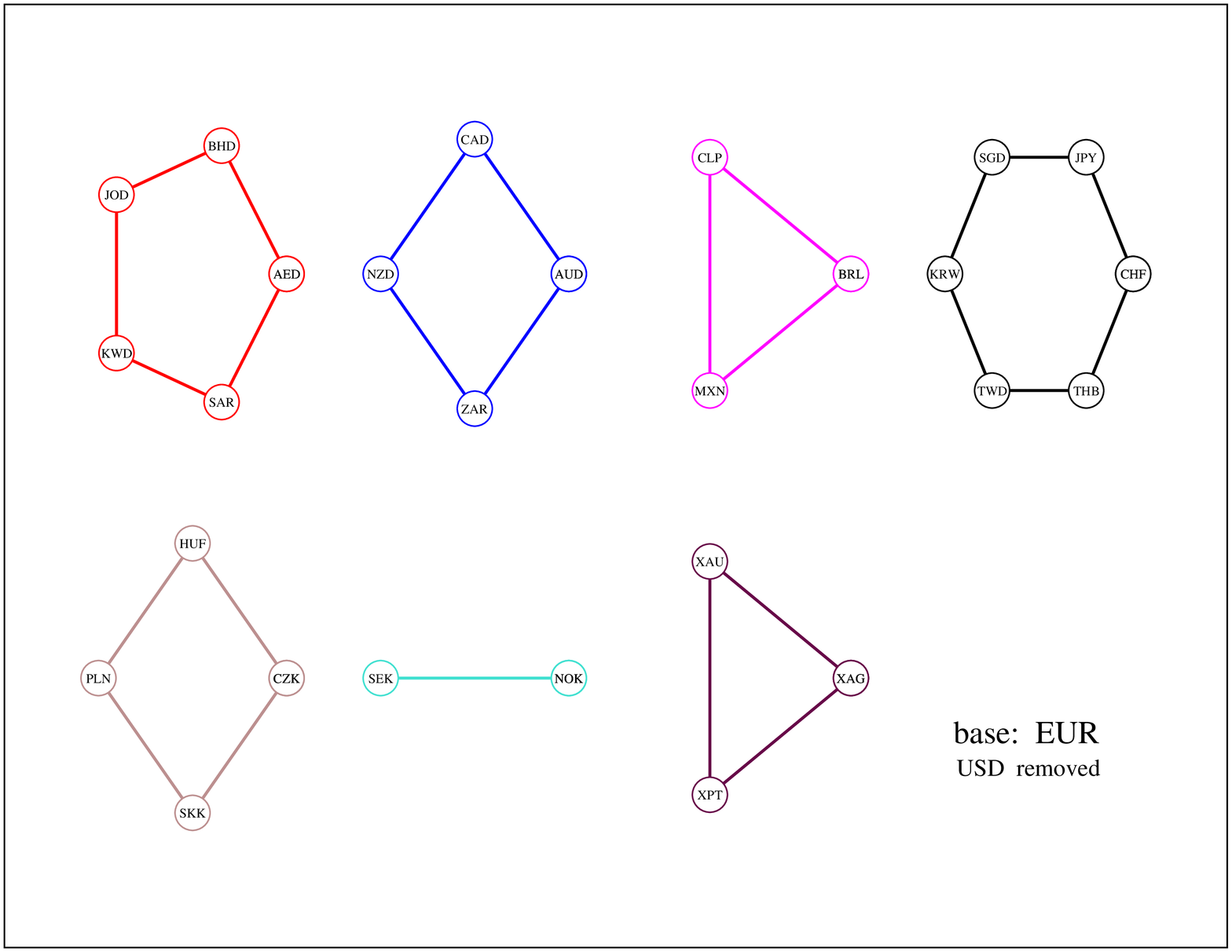}
\hspace{-0.1cm}
\epsfxsize 6cm
\epsffile{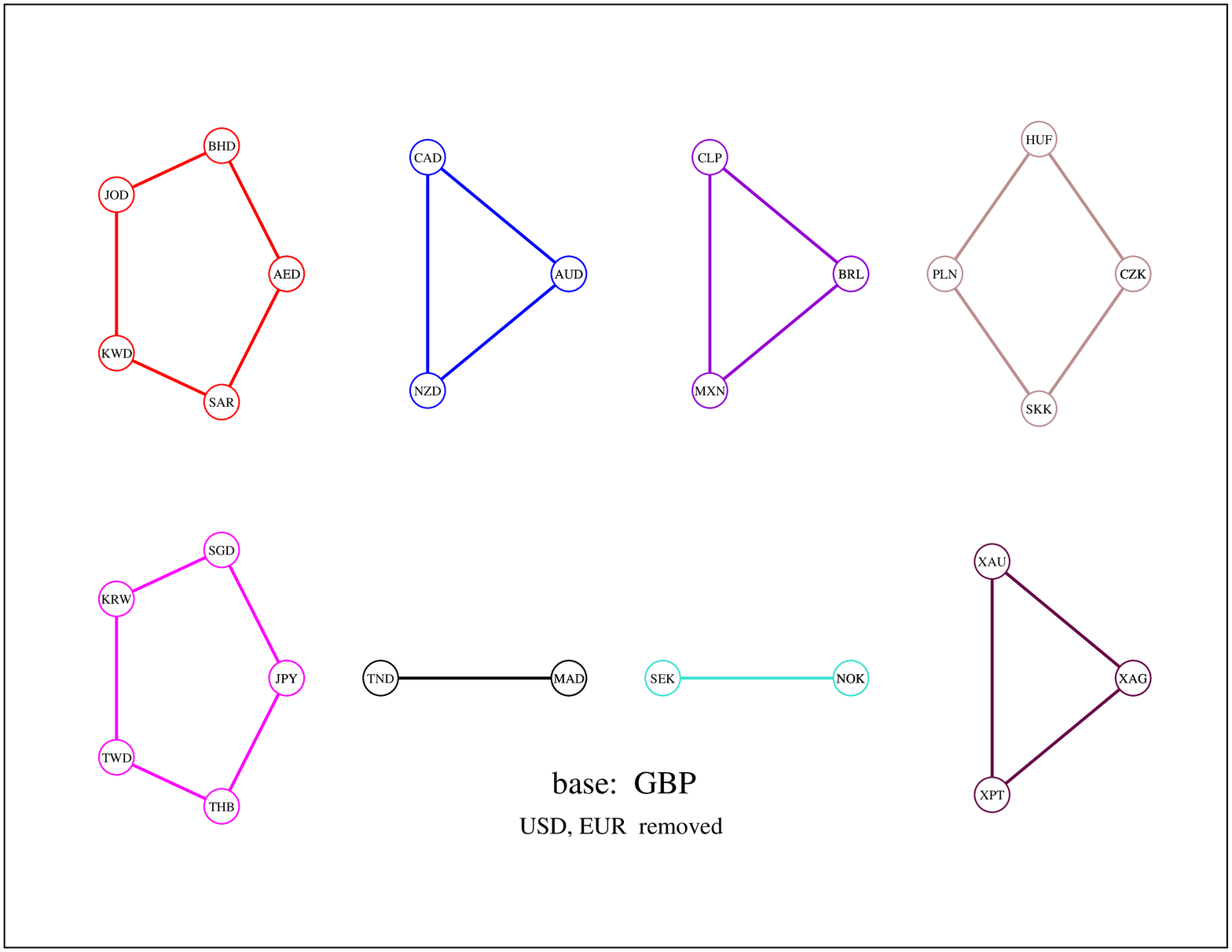}

\vspace{0.0cm}
\hspace{0.1cm}
\epsfxsize 6cm
\epsffile{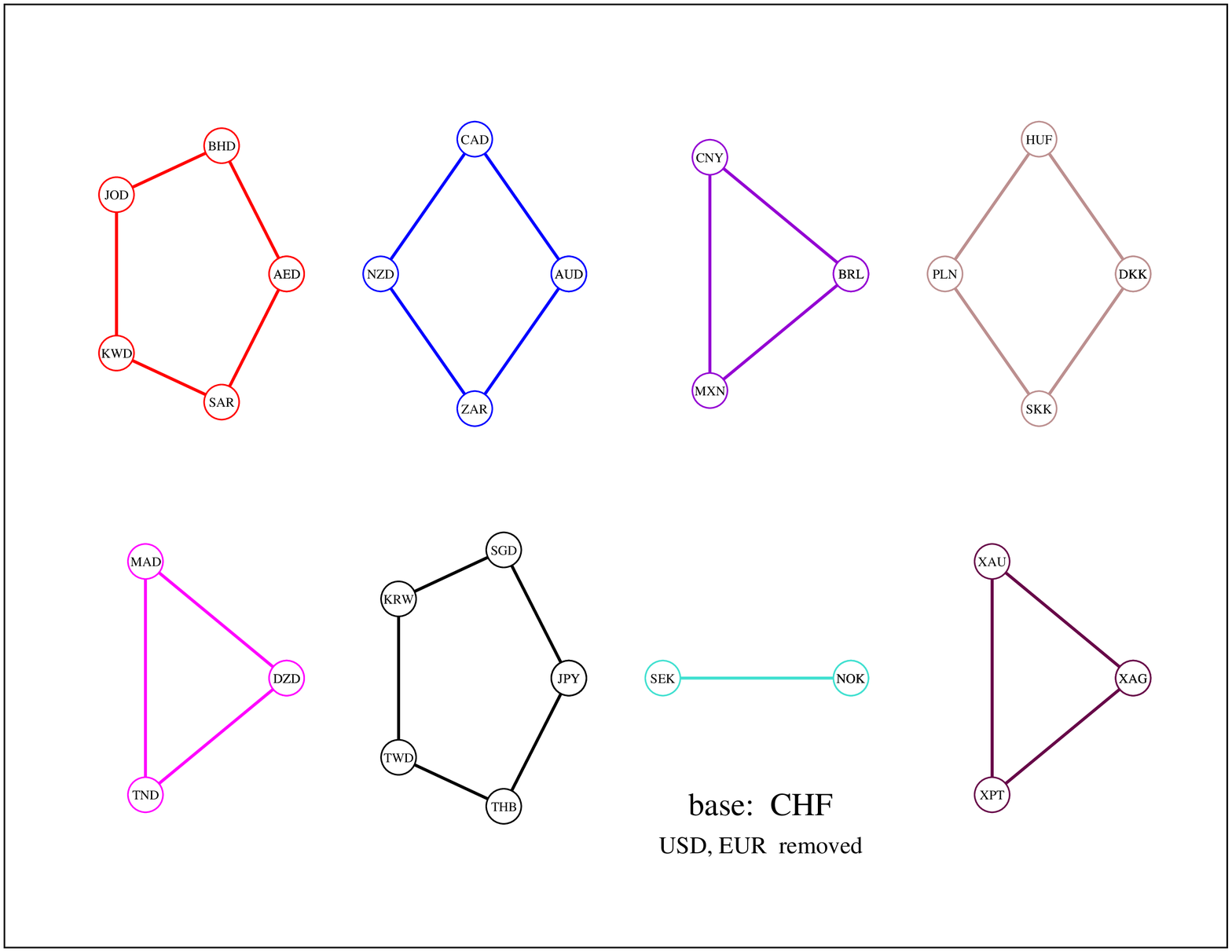}
\hspace{-0.1cm}
\epsfxsize 6cm
\epsffile{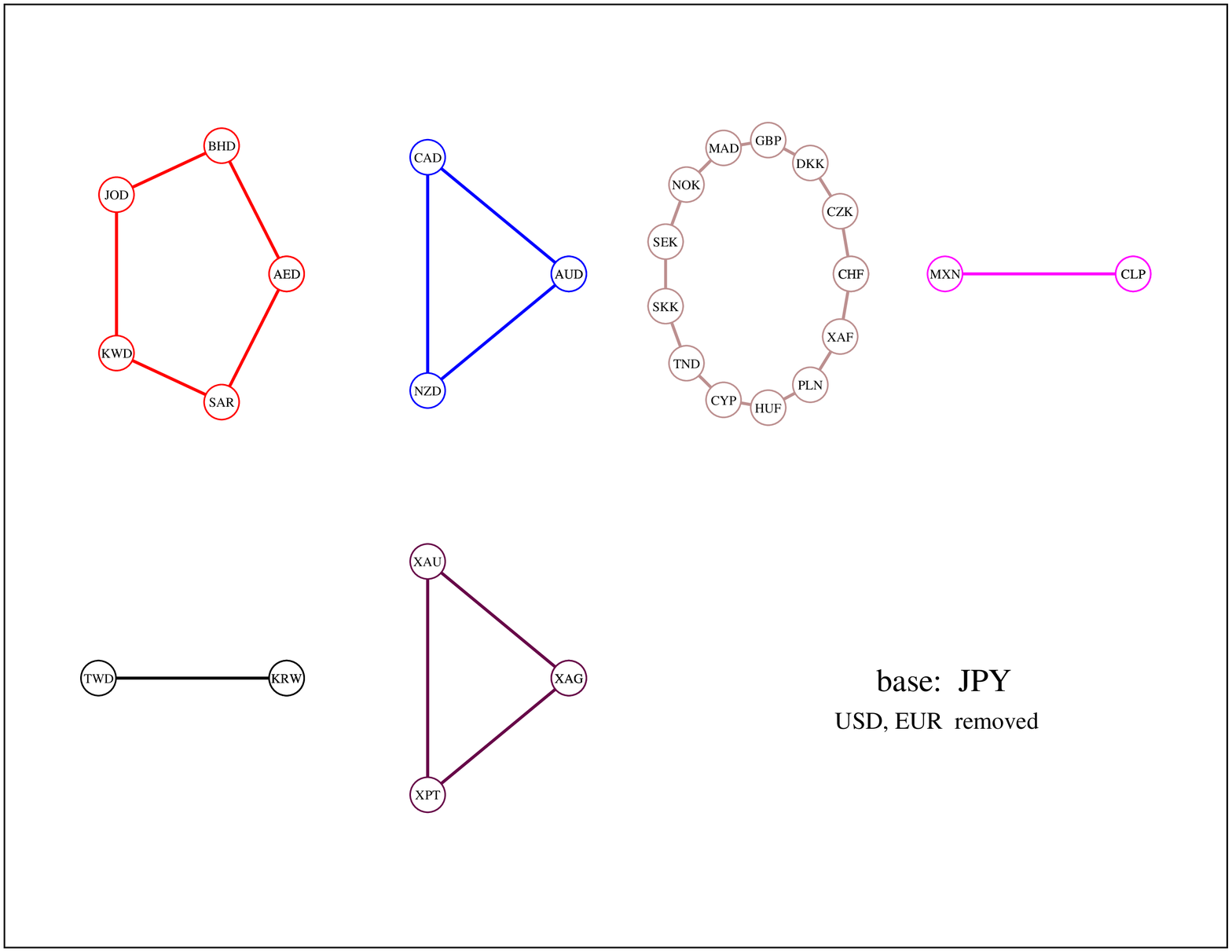}

\vspace{0.0cm}
\hspace{0.1cm}
\epsfxsize 6cm
\epsffile{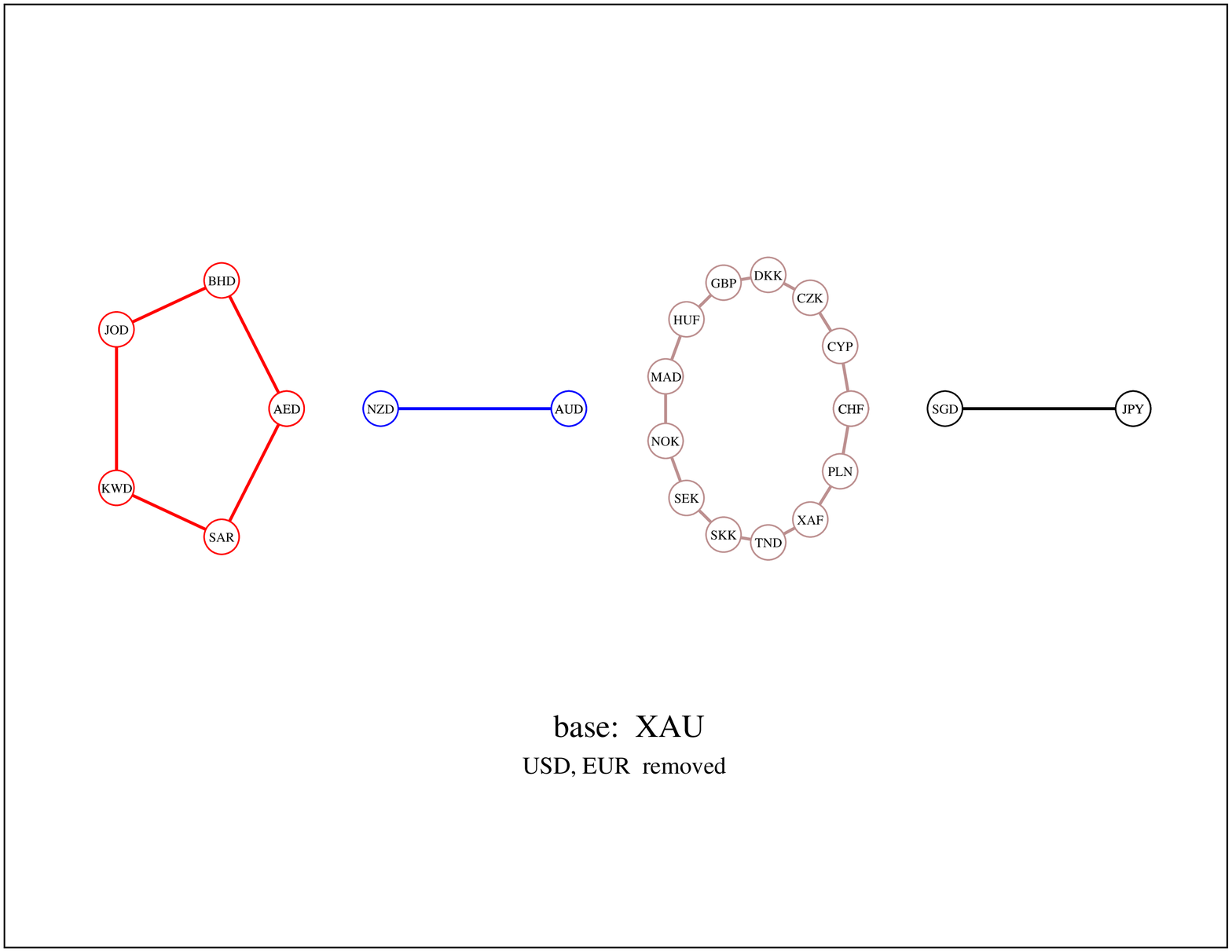}
\hspace{-0.1cm}
\epsfxsize 6cm
\epsffile{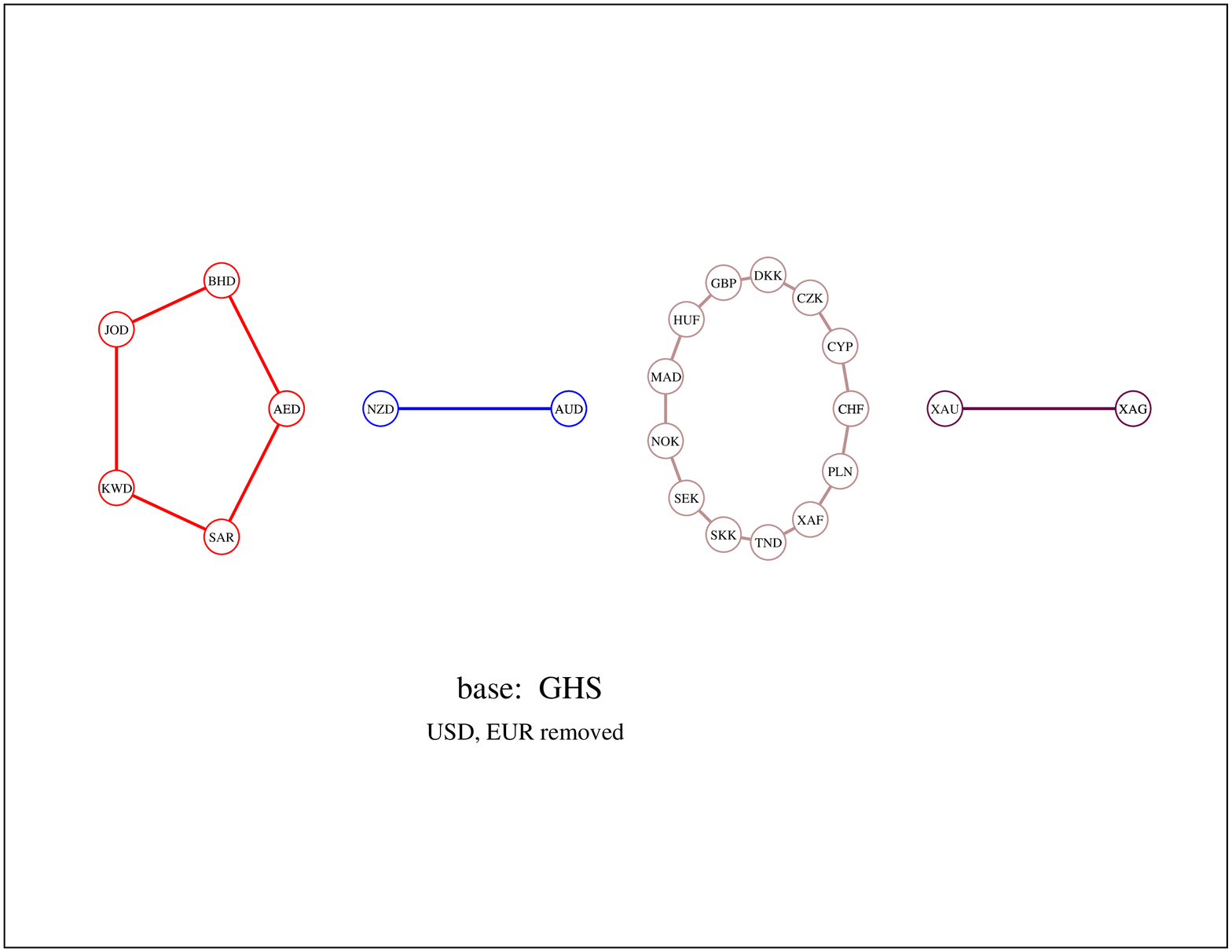}
\vspace{0.2cm}
\caption{Clusters of currencies identified in the network of exchange
rates B/X for six exemplary choices of the base currency B, after removing
contributions of the B/USD and B/EUR cross-rates. Specific colors denote  
clusters associated with different geographical regions. Note that cluster
schemes do not show all links between the nodes.}
\end{figure}

Figure 4 shows the clusters that have been found for the same six base 
currencies as in Figures 1 and 3 (in all Figures we label nodes, i.e., the 
exchange rates B/X, only with the term currency X, dropping the base B, 
since B is common to all exchange rates in a particular network 
representation). The cluster structure of the FX network is considerably 
stable. There are clusters, like the Middle East cluster 
AED-BHD-JOD-KWD-SAR and the commodity-trade-related cluster 
AUD-NZD-(CAD)-(ZAR), which are present in all network representations, 
there are also ones which can be found only in some representations 
(e.g. the Central European cluster CZK-HUF-PLN-SKK, the Scandinavian 
cluster NOK-SEK, and the precious metals cluster XAG-XAU-XPT). In general, 
among the analyzed network representations there are two dominating 
patterns of the cluster structure: the first one for CHF, EUR, GBP and 
USD, and the second one for JPY, XAU and GHS. This overlaps with the two 
different patterns of the eigenvalue spectra shown in Figure 3. Our 
results show that the currencies group together primarily according to 
geographical factors, but sometimes also according to other factors like, 
for example, the commodity trade.

\subsection{Minimal spanning tree}

A useful tool in analysis of the structure of a network is the minimal 
spanning tree method (MST), which allows one to describe and show the most 
important features of the network structure graphically in a compact 
form. For example, the B-based currency network is a fully connected 
undirected network with $N - 1$ nodes and $(N-1)(N-2)/2$ connections;
the minimal spanning tree graph in this case has the same number of nodes 
but only $N - 2$ connections.

\begin{figure}
\hspace{1.0cm}
\epsfxsize 10cm
\epsffile{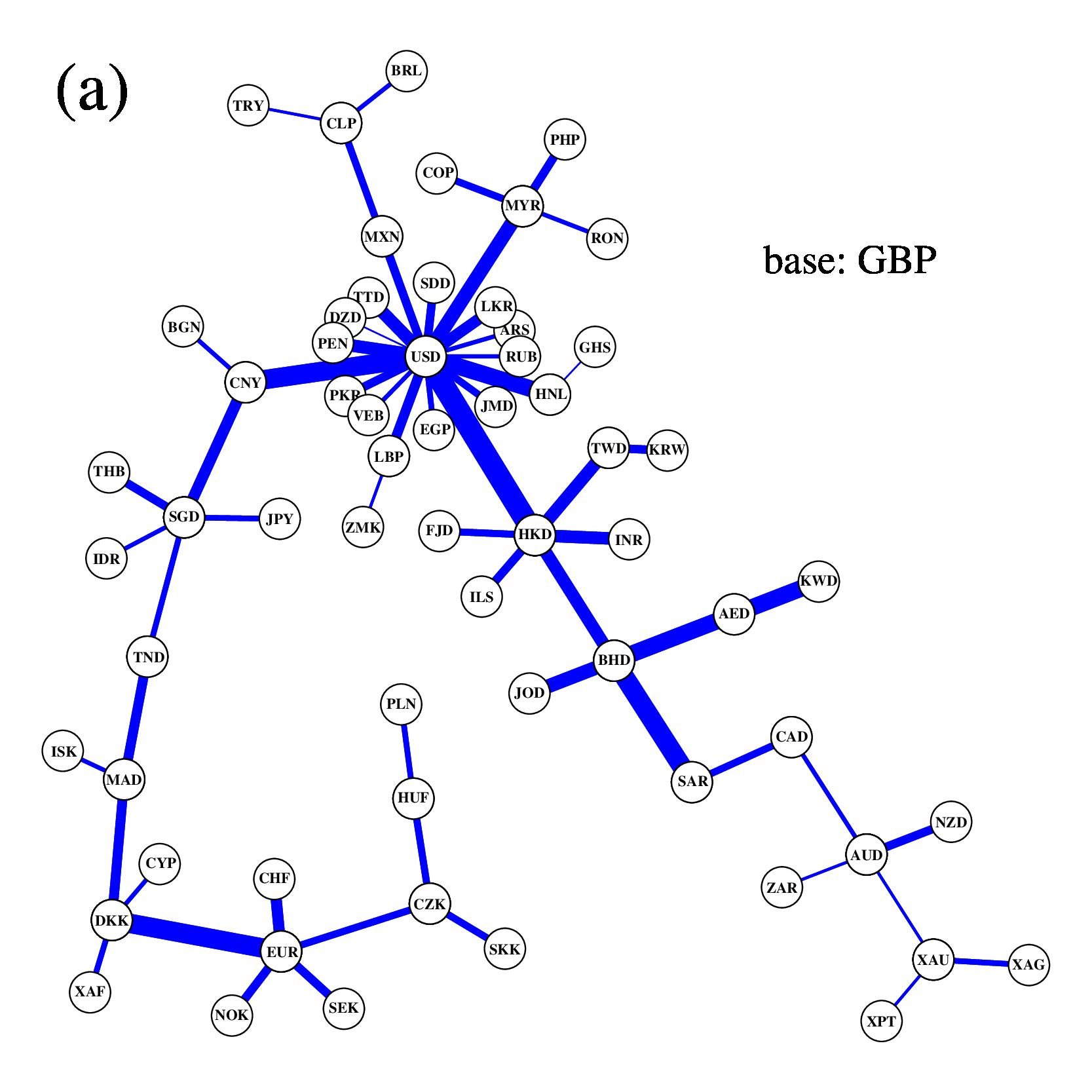}

\vspace{0.0cm}
\hspace{1.0cm}
\epsfxsize 10cm 
\epsffile{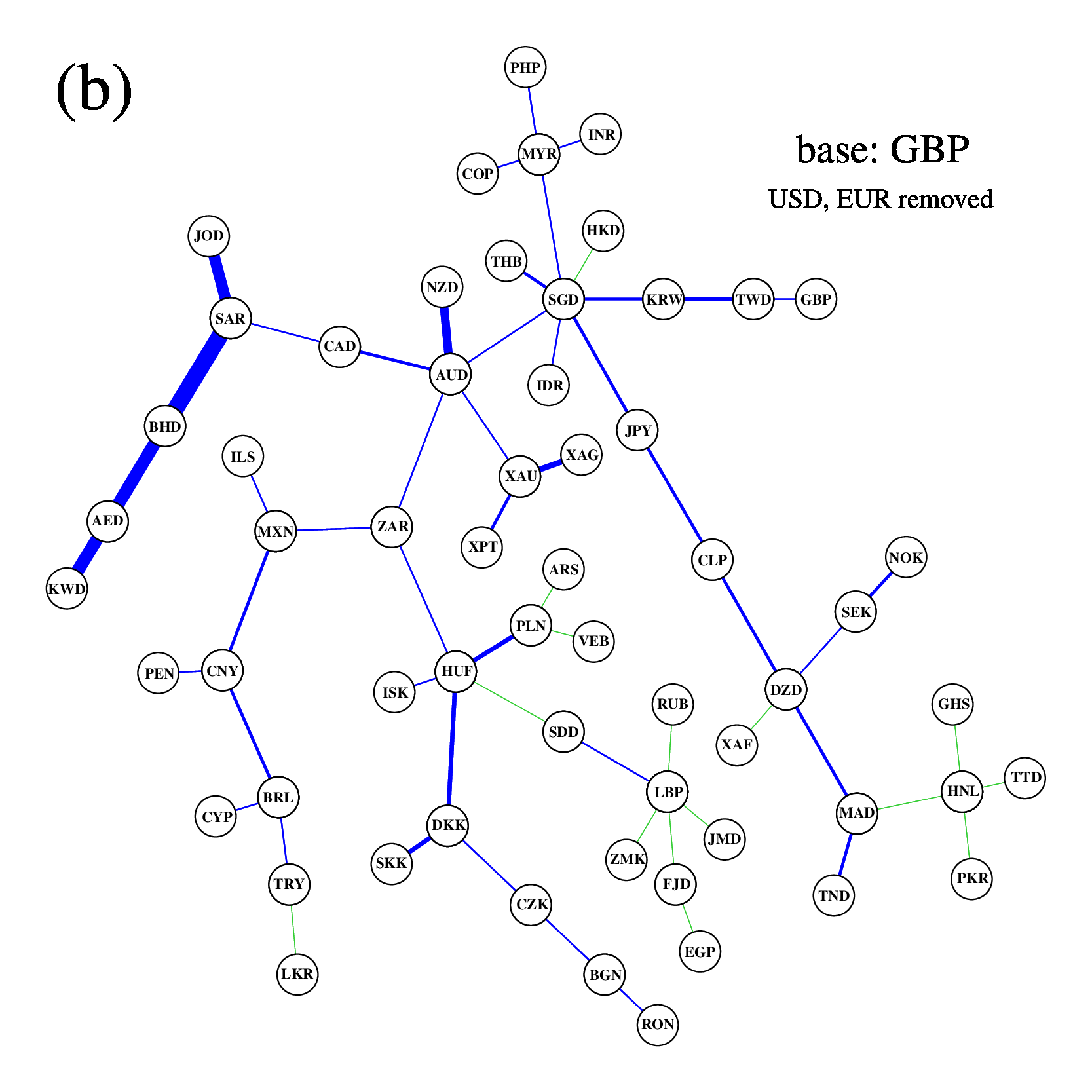}
\caption{Minimal spanning trees for the GBP-based network of exchange
rates for the original data (a) and for the data after removing
contribution of the GBP/USD and GBP/EUR exchange rates (b). Entire period
1999-2008 is considered. Line widths are proportional to the correlation
coefficients for the corresponding exchange rates. Anticorrelated nodes   
are denoted by green edges.}
\end{figure}

The method is based on a metric $d_{\rm X,Y}^{\rm B}$ defined on the 
entries of the correlation matrix by the formula:
\begin{equation}
d_{\rm X,Y}^B = \sqrt{2(1 - C_{\rm X,Y}^{\rm B})}.
\end{equation}
In our case this quantity measures the distance between two exchange rates 
B/X and B/Y. For completely correlated signals $d_{\rm X,Y}^{\rm B} = 0$ 
and for completely anticorrelated ones $d_{\rm X,Y}^{\rm B} = 2$. MST is 
then constructed by sorting the list of distances calculated for all pairs 
(X,Y) and by connecting the closest nodes with respect to $d_{\rm 
X,Y}^{\rm B}$ in such a manner that each pair of nodes is connected 
exactly via one path. Each edge represents a link between this node and
its closest neighbour. Detailed instructions can be found e.g. 
in~\cite{mantegna99}. MST assigns to each node a measure of its importance 
in a hierarchy of nodes: a node is more significant if it has a higher 
degree $K$, i.e. has more edges attached to it or, in a case of weighted 
networks, has edges with high weights.

\begin{figure}
\hspace{1.0cm}
\epsfxsize 10cm
\epsffile{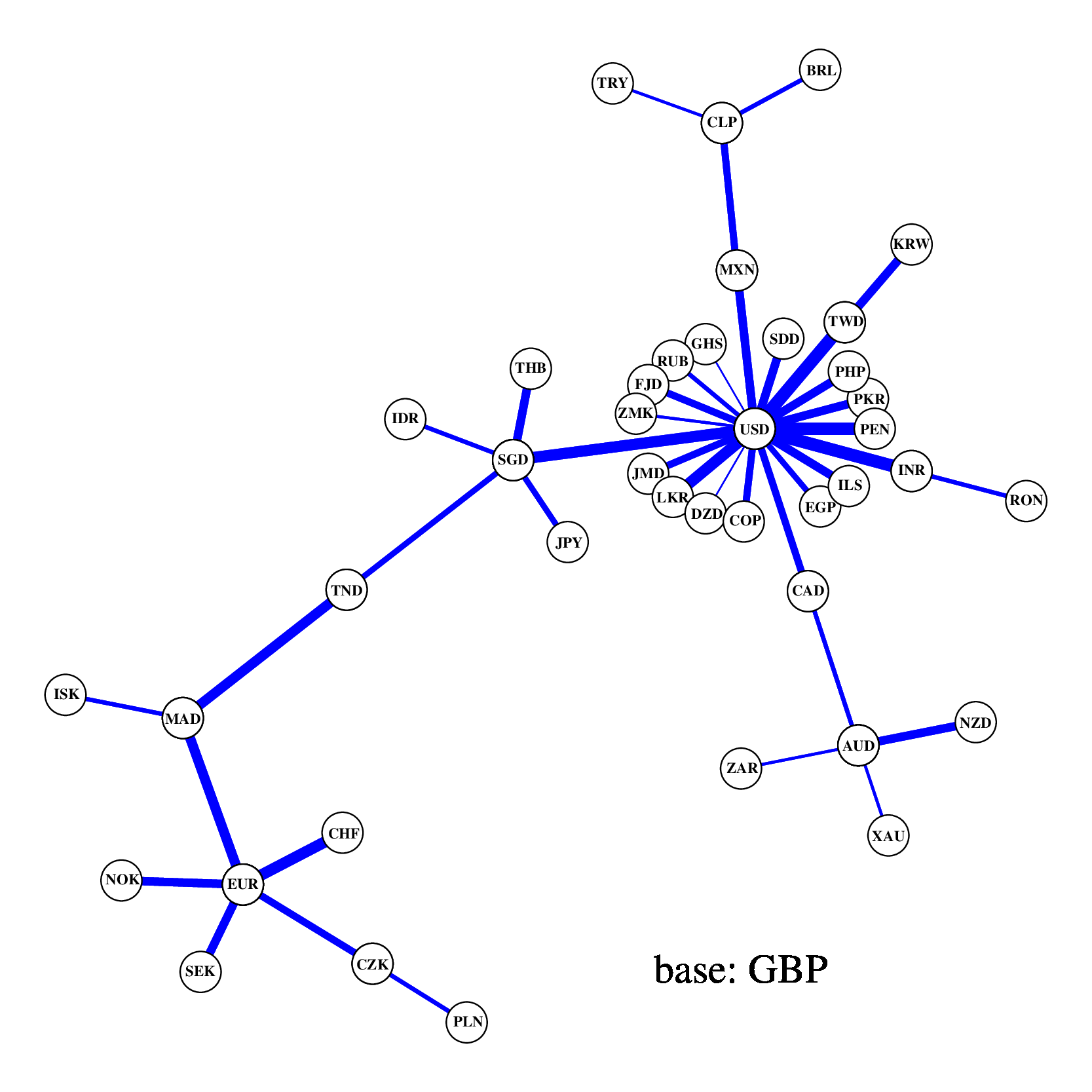}
\caption{Minimal spanning tree for the GBP-based network of 41 exchange
rates GBP/X for a set of independent currencies X. Line widths are
proportional to the correlation coefficients for the corresponding pairs
of the exchange rates.}
\end{figure}

The complete minimal spanning tree for the GBP-based network is plotted in 
Figure 5(a). This choice of B allows us to observe the most complete 
cluster structure (see Figure 4(b)). In agreement with our remark from the 
previous section, the most important node is the node related to USD with 
a degree $K = 17$ (it is directly linked to 28\% of nodes). Other 
important nodes are HKD ($K=6$), SGD and EUR ($K=5$), as well as BHD, AUD, 
DKK and MYR ($K=4$). Some edges are particularly strong (heavy lines in 
Figure 5) which typically indicates that one of the associated currencies 
is artificially pegged to the other. This is the case, for instance, of 
DKK-EUR, HKD-USD, MYR-USD and so on. Pegs lead to a situation in which 
certain nodes, in fact primarily coupled to EUR or USD, are effectively 
connected to less important currencies as DKK or HKD. This is why HKD has 
a larger multiplicity in Figure 5(a) than EUR, and a few other nodes have 
significant degrees even if they do not belong to the group of major 
currencies (BHD is a striking example here). Without this effect both USD 
and EUR would have a much larger value of $K$.

In order to avoid this spurious phenomenon of absorbing a fraction of 
centrality of the major currencies by their satellites, we single out only 
those exchange rates GBP/X which involve independent currencies, i.e. such 
currencies which, in the analyzed interval of time, were not explicitely 
pegged to other monetary units. Figure 6 shows the corresponding MST 
comprising 41 nodes. Now the tree looks different. The USD node is even 
more central ($K=19$, direct links with 40\% of nodes) than for the full 
set of currencies in Figure 5(a), and it is followed by EUR ($K=5$), SGD
($K=5$) and AUD ($K=4$). This better reflects the role played by 
USD, EUR and AUD in the world's currency system. SGD, which is here the 
least important unit, owes its localization in the MST to its specific 
basket-oriented regulation.

\begin{figure}
\hspace{0.5cm}
\epsfxsize 12cm
\epsffile{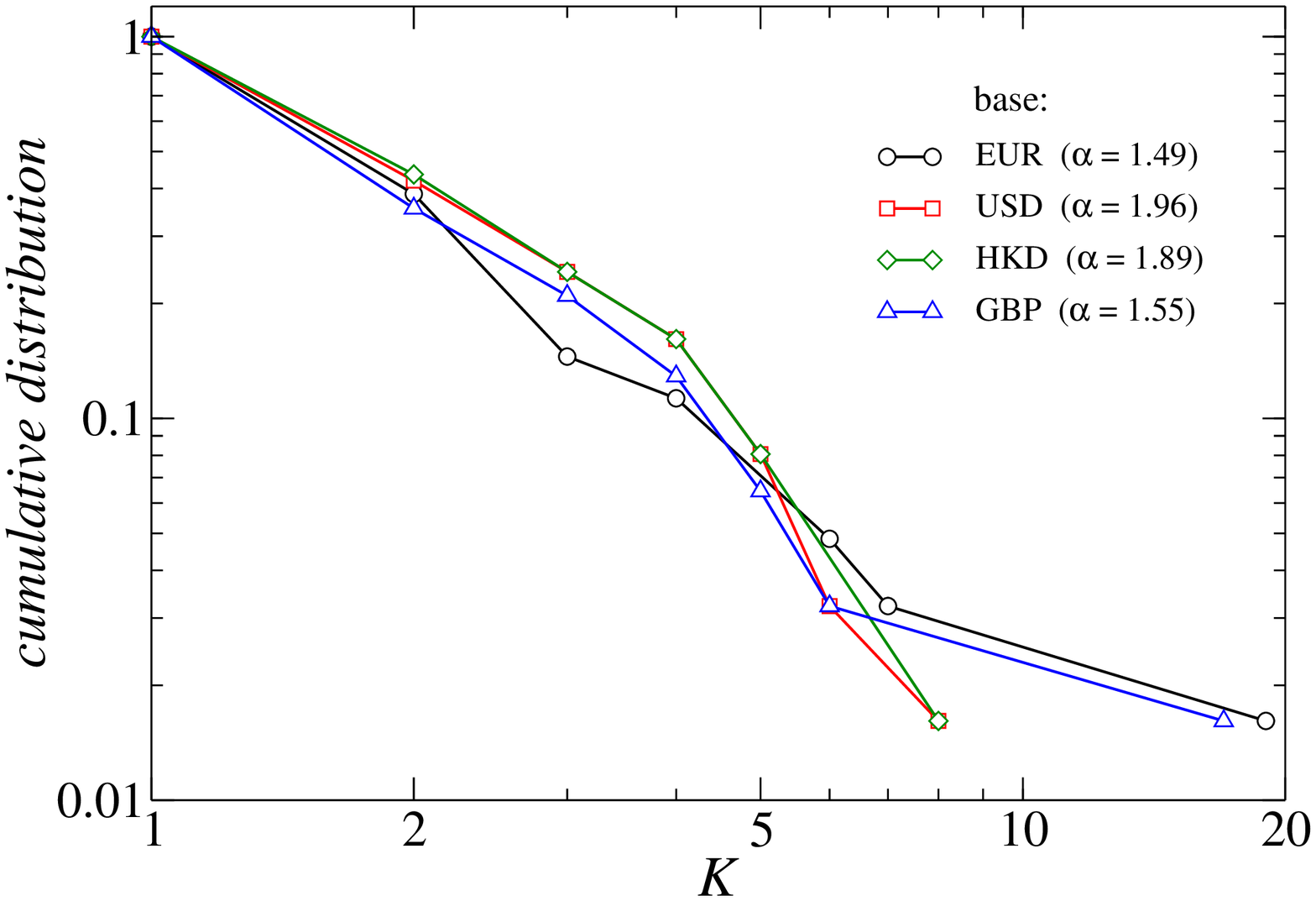}

\vspace{-0.5cm}
\hspace{0.5cm}
\epsfxsize 12cm
\epsffile{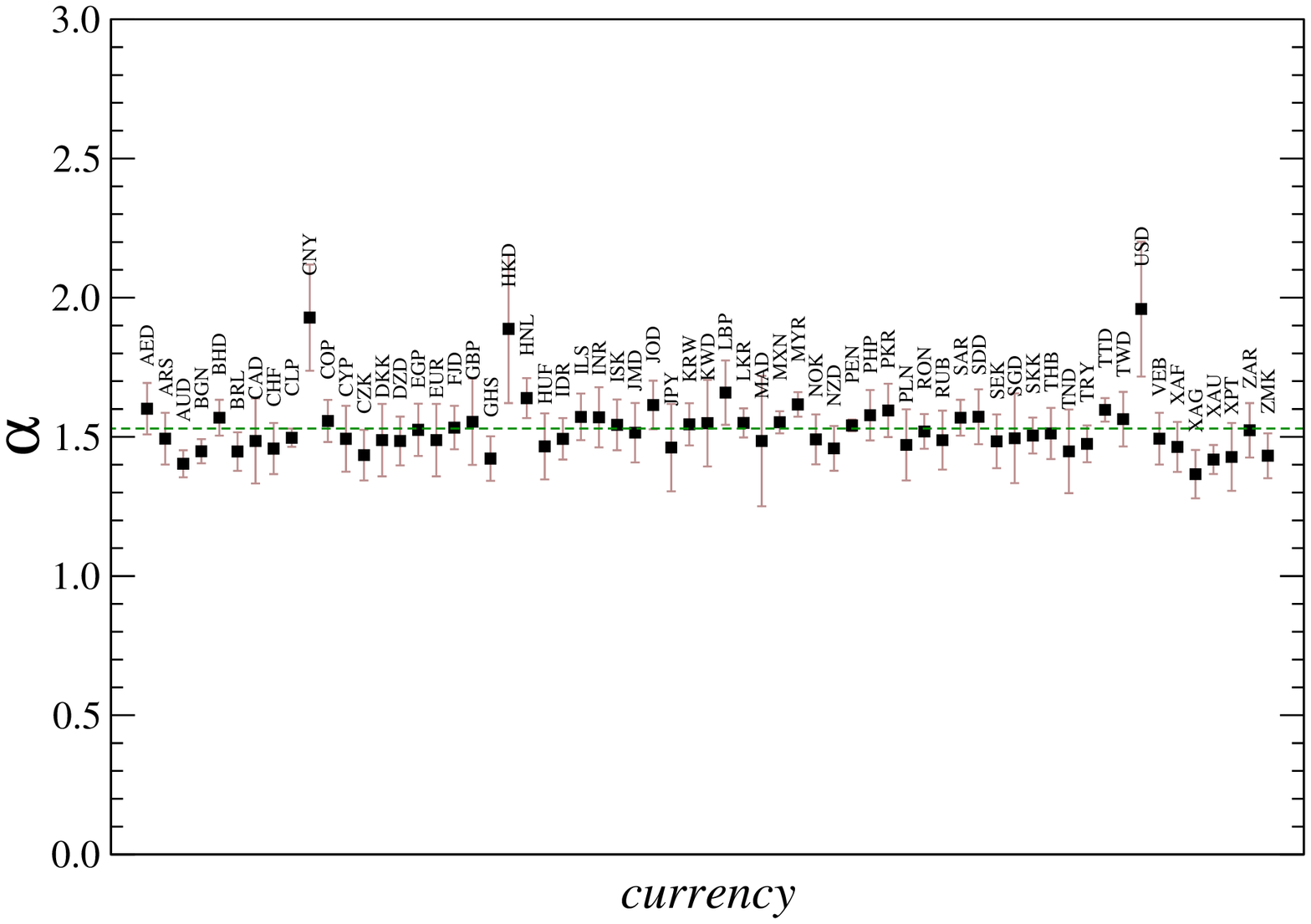}
\caption{(Top) Cumulative distributions of the MST node degrees for 
exemplary representations of the currency network together with the fitted 
values of scaling exponent $\alpha$. (Bottom) Scaling exponents $\alpha$ 
fitted to the cumulative distributions of the node degrees for all 63 
representations of the network.}
\end{figure}

Figures 5(a) and 6 show that the currency network has a hierarchical 
structure, which supports earlier 
findings~\cite{mizuno06,naylor07,gorski08}. These graphs, however, do not 
allow to observe the clustered structure discussed in the previous 
section. However, by neglecting the influence of the GBP/USD and GBP/EUR 
exchange rates (Eq.(\ref{removal})), most clusters can easily be revealed. 
Indeed, in Figure 5(b) the cluster of AUD-CAD-NZD, as well as the ones of 
the Maghreb currencies, the Central European currencies, the Middle East 
currencies, the South-East Asian currencies, and the precious metals are 
identifiable. The hierarchical structure of the original network is lost 
here, however: the incomplete network of Figure 5(b) resembles rather a 
random graph.

In order to be able to say something more on the MST topology, we 
calculate a distribution of the node degrees for the GBP-based network 
from Figure 5(a). We count the number of nodes of the same degree and 
calculate the cumulative distribution function for this quantity. This 
empirical distribution resembles the scale-free power-law behaviour, thus 
we attempt to fit the power function and evaluate the scaling exponent 
$\alpha$. For the GBP-based MST $\alpha = 1.55 \pm 0.16$.

\begin{figure}
\hspace{1.0cm}
\epsfxsize 10cm
\epsffile{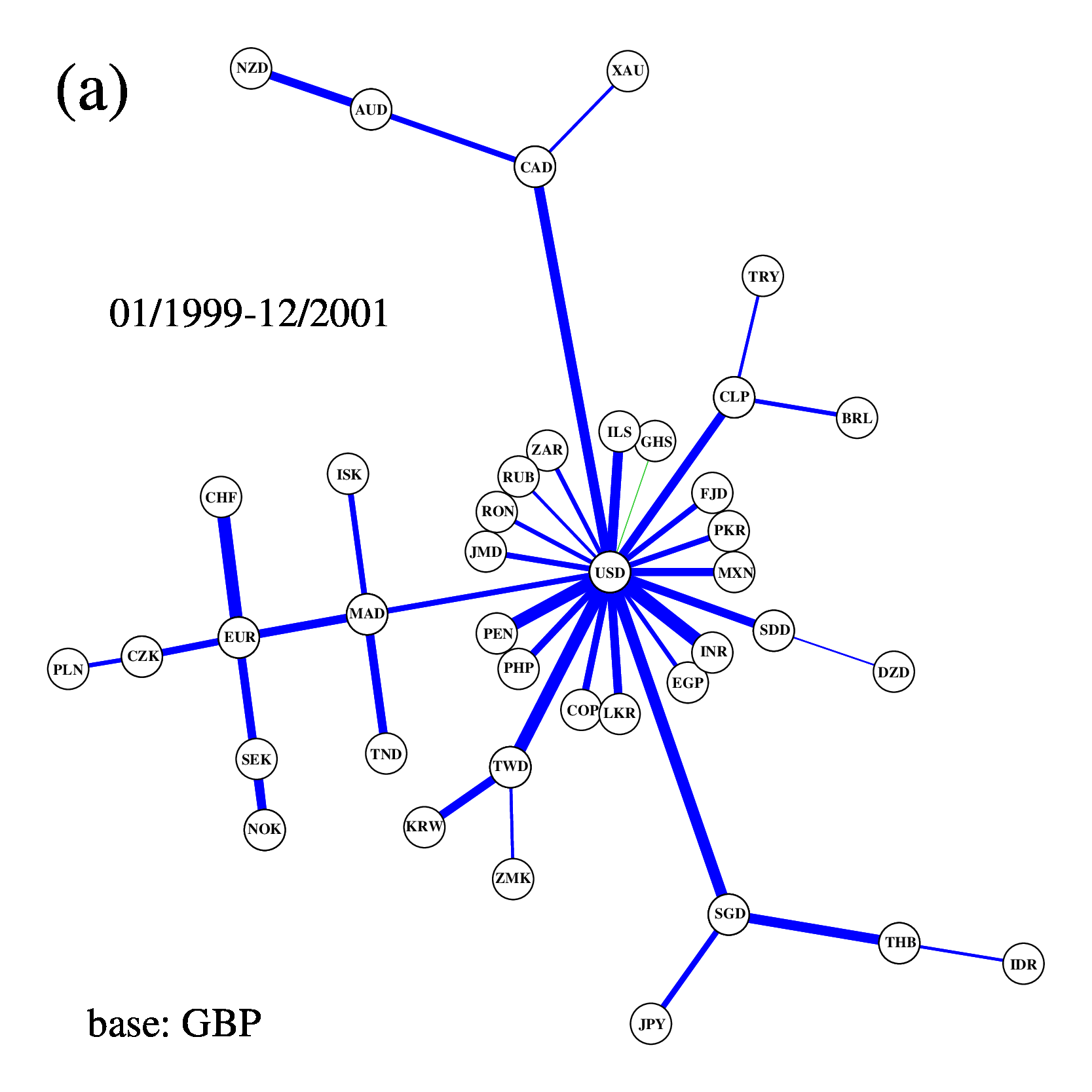}

\vspace{0.0cm}
\hspace{1.0cm}
\epsfxsize 10cm
\epsffile{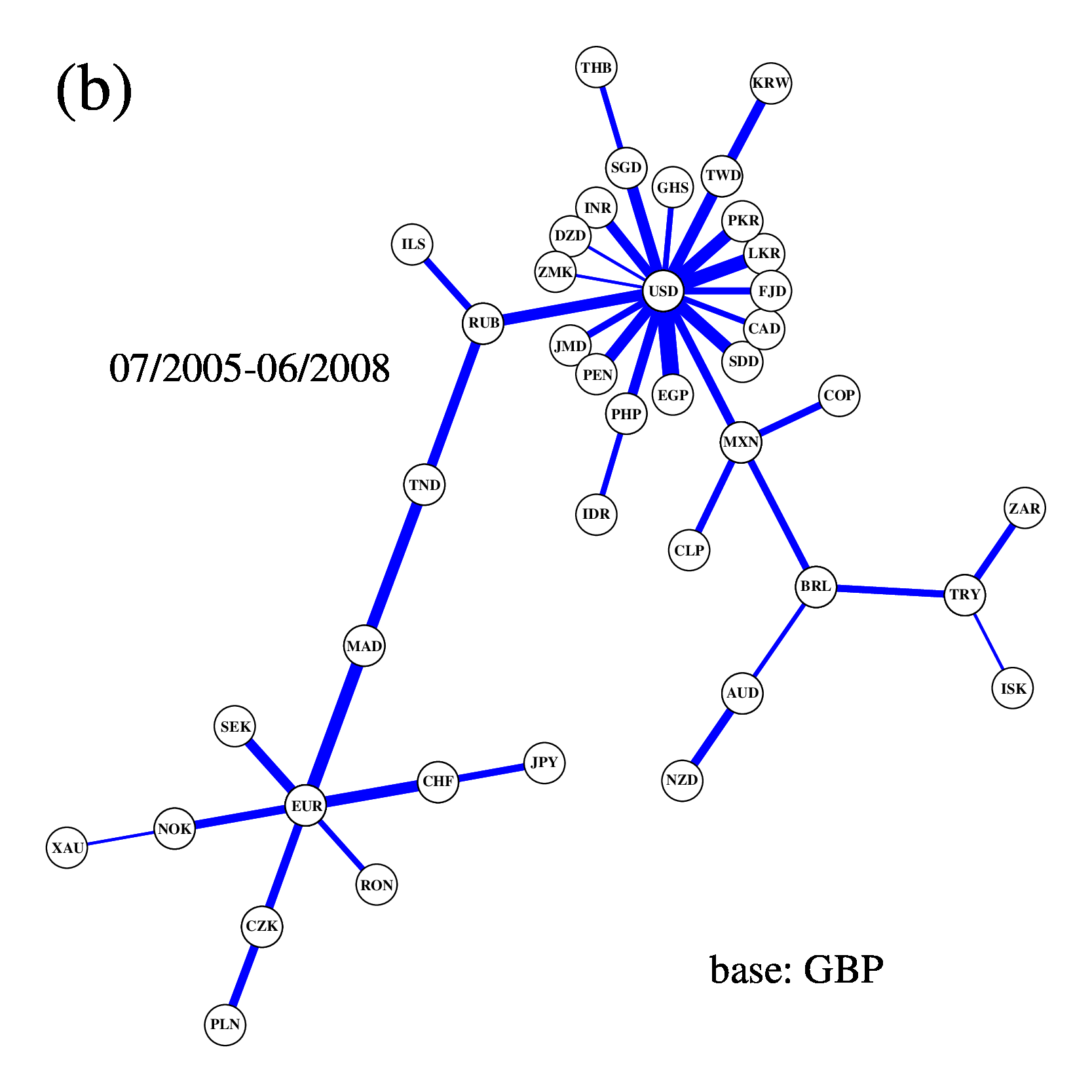}
\caption{Minimal spanning trees for the GBP-based network for two 
different time intervals: the first three years (a) and the last three 
years (b) of the period spanned by our data. Line widths are proportional 
to the correlation coefficients for the corresponding pairs of the 
exchange rates.}
\end{figure}

It is worthwhile to compare the scaling exponents for the cumulative
distributions of the node degrees using different base currencies.  
Typically, the scaling relations exhibited by the c.d.f.s are of
statistical significance and the corresponding scaling exponents can be
estimated with only a small error~\cite{gorski08}. The exponents have
values in the range $1.37 \le \alpha \le 1.96$ but a vast majority of
values do not exceed 1.66. Only for five base currencies $\alpha > 1.66$:  
this happens for USD ($1.96 \pm 0.24$), CNY ($1.93 \pm 0.19$), and HKD
($1.89 \pm 0.27$). In all these cases the scaling quality is also poor, as
the significant statistical errors indicate. Both HKD and CNY are tied to
USD and therefore they mimic its evolution. Moreover, they develop a small
collective eigenvalue $\lambda_1^{\rm B}$ (Figure 2), which is also a
property inherited from the US dollar. This result suggests that the base
currencies with high values of $\alpha$ are associated with the network
representations that are more random than typical hierarchical networks.
It occurs that the exponent averaged over all base currencies
$\bar{\alpha} = 1.53 \pm 0.11$, in agreement with the theoretically 
derived value of 1.61 for a hierarchical network of the same 
size~\cite{ravasz03,noh03}. The above results are displayed in Figure 7.

\subsection{Temporal stability of the market structure}

\begin{figure}
\hspace{1.0cm}
\epsfxsize 10cm
\epsffile{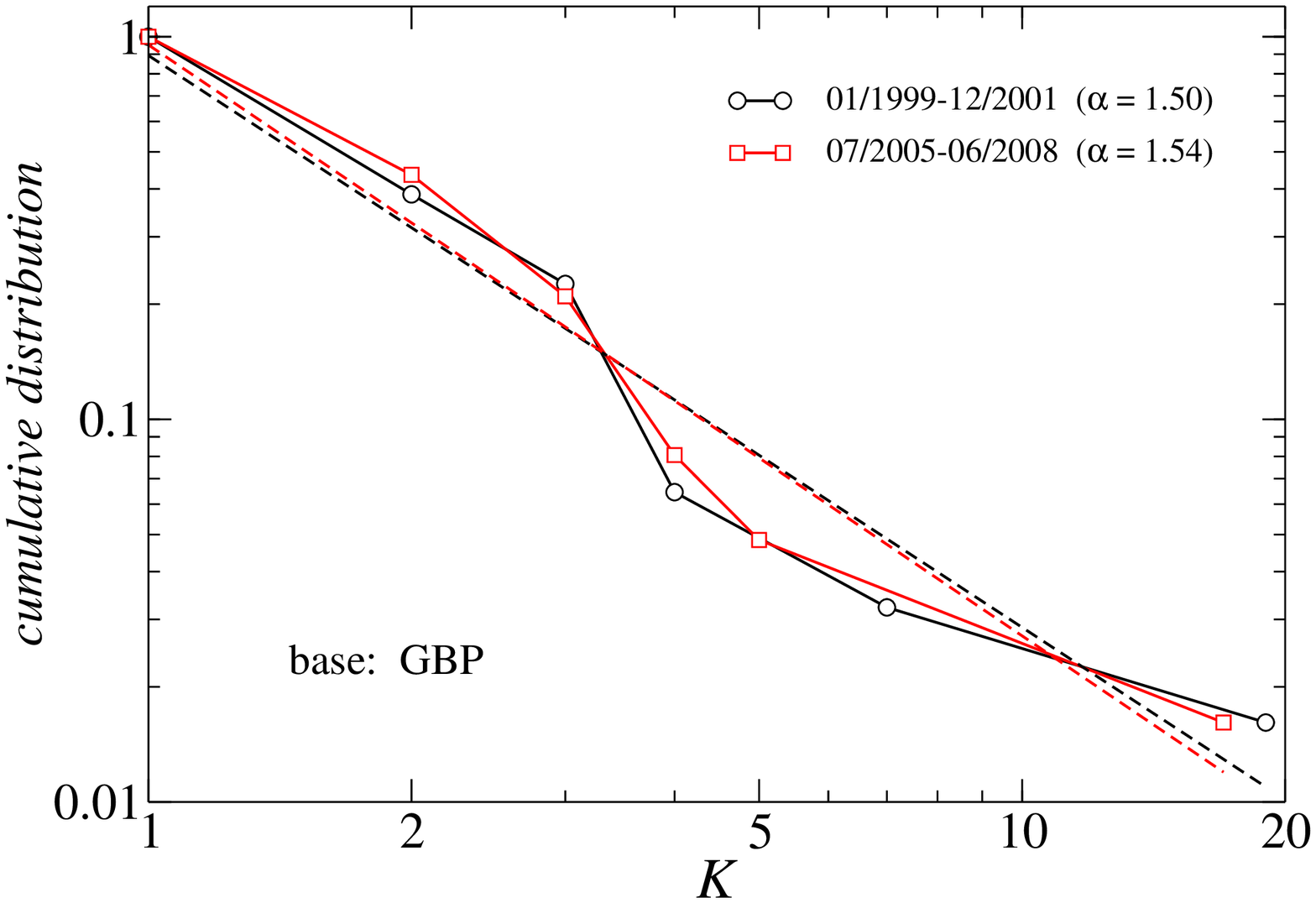}

\vspace{-0.5cm}
\hspace{1.0cm}
\epsfxsize 10cm
\epsffile{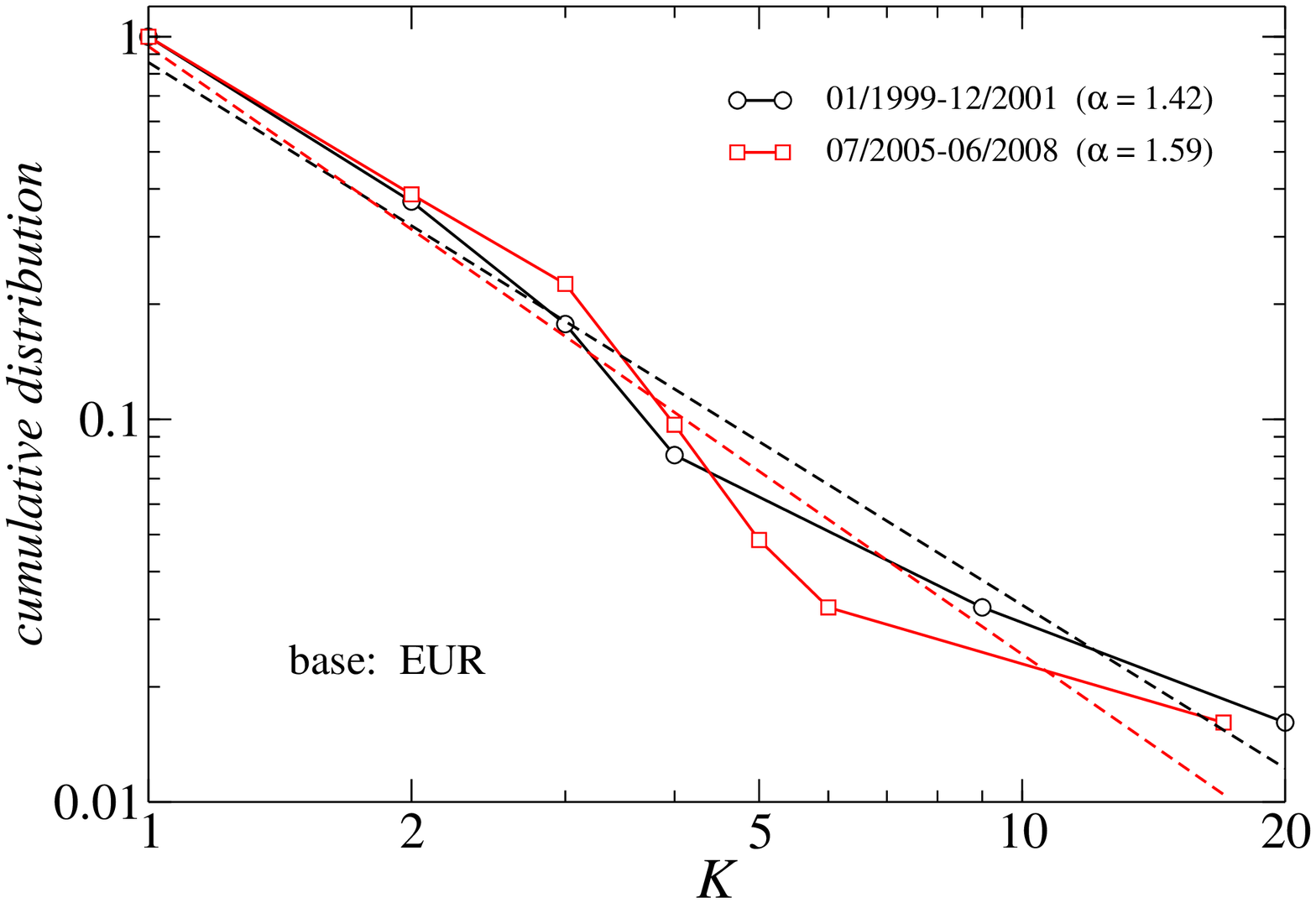}
\vspace{0.3cm}
\caption{Cumulative distributions of node degrees $K$ for the GBP-based
(top) and the EUR-based (bottom) minimal spanning trees for two 
three-years-long time intervals: 01/1999-12/2001 (black) and 
07/2005-06/2008 (red). An attempt of fitting power functions to these 
distributions is also presented (denoted by dashed lines).}
\end{figure}

Earlier works showed that the currency networks, expressed by the MST
graphs, are sensitive to current market situation - for instance, to which
currencies are most active at the moment~\cite{mcdonald05,naylor07}.  
However, despite this fact a majority of network edges were reported 
to be rather stable.

\begin{figure}
\epsfxsize 13cm
\epsffile{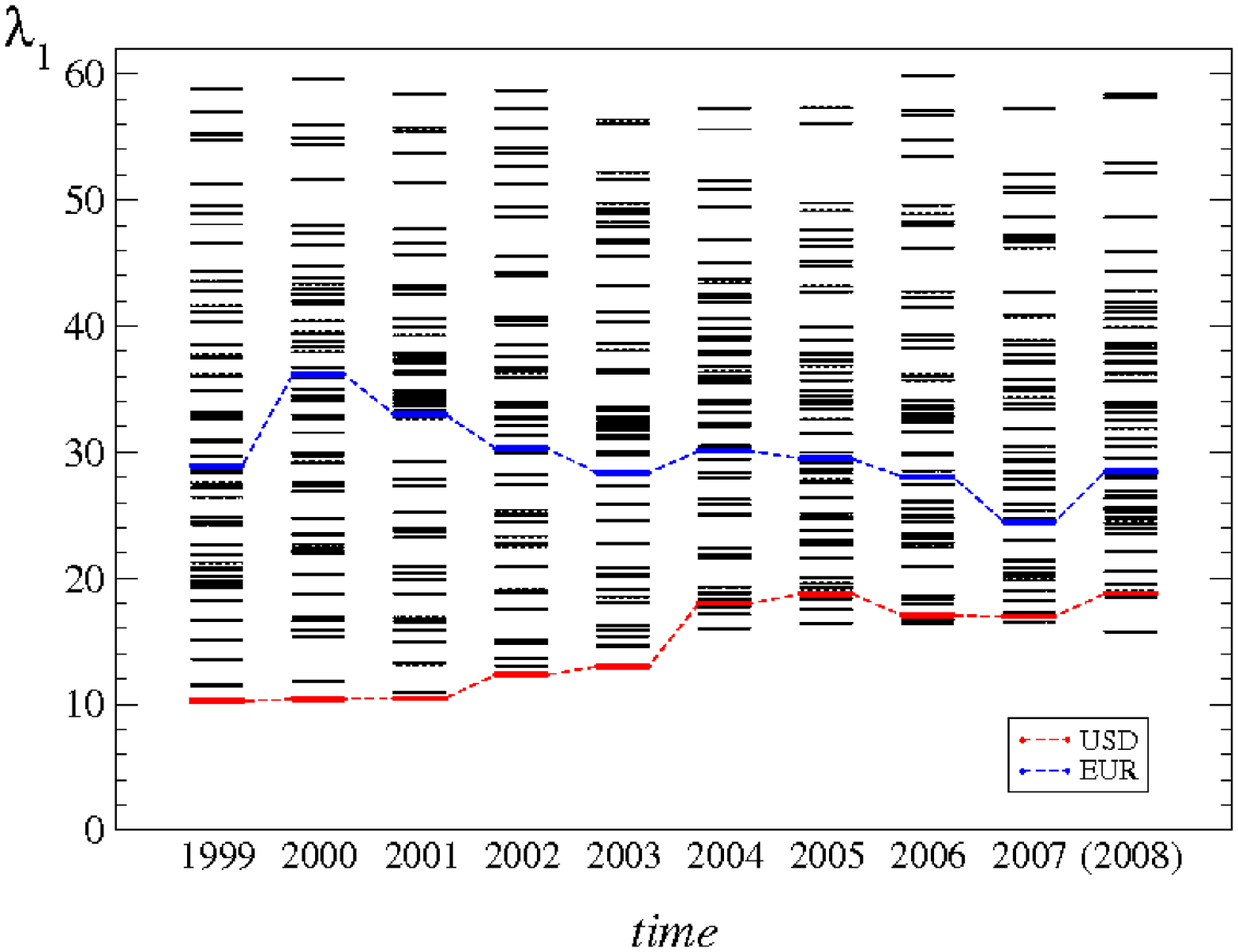}
\caption{Annual changes of the largest eigenvalue $\lambda_1^{\rm B}$
ladder. Positions of $\lambda_1^{\rm USD}(t)$ and $\lambda_1^{\rm EUR}(t)$
are distinguished and connected by dashed lines.}
\end{figure}

As an example of the MST variability, in Figure 8 we present the exemplary
GBP-based trees for the two disjoint and mutually distant time intervals
of three years: 01/1999-12/2001 (Interval 1) and 07/2005-06/2008 (Interval
2). To eliminate the spurious wandering of nodes between the major
currencies and their satellites, we restricted the network to 41
independent exchange rates (as it was the case for the entire period in
Figure 6). In fact, the structure of MSTs in both panels of Figure 5 is
different. In the more recent Interval 2 (Figure 5(b)) the USD node has
smaller centrality than in the earlier Interval 1 (Figure 5(a)); its
degree dropped from $K=21$ to $K=17$. In addition, the tree in Figure 5(b)
has more extended branches than its counterpart in Figure 5(a). However,
neither of the trees lacks the overall hierarchical structure, which seems
to be a stable property of the network under study.

\begin{figure}
\hspace{-0.2cm}
\epsfxsize 7cm
\epsffile{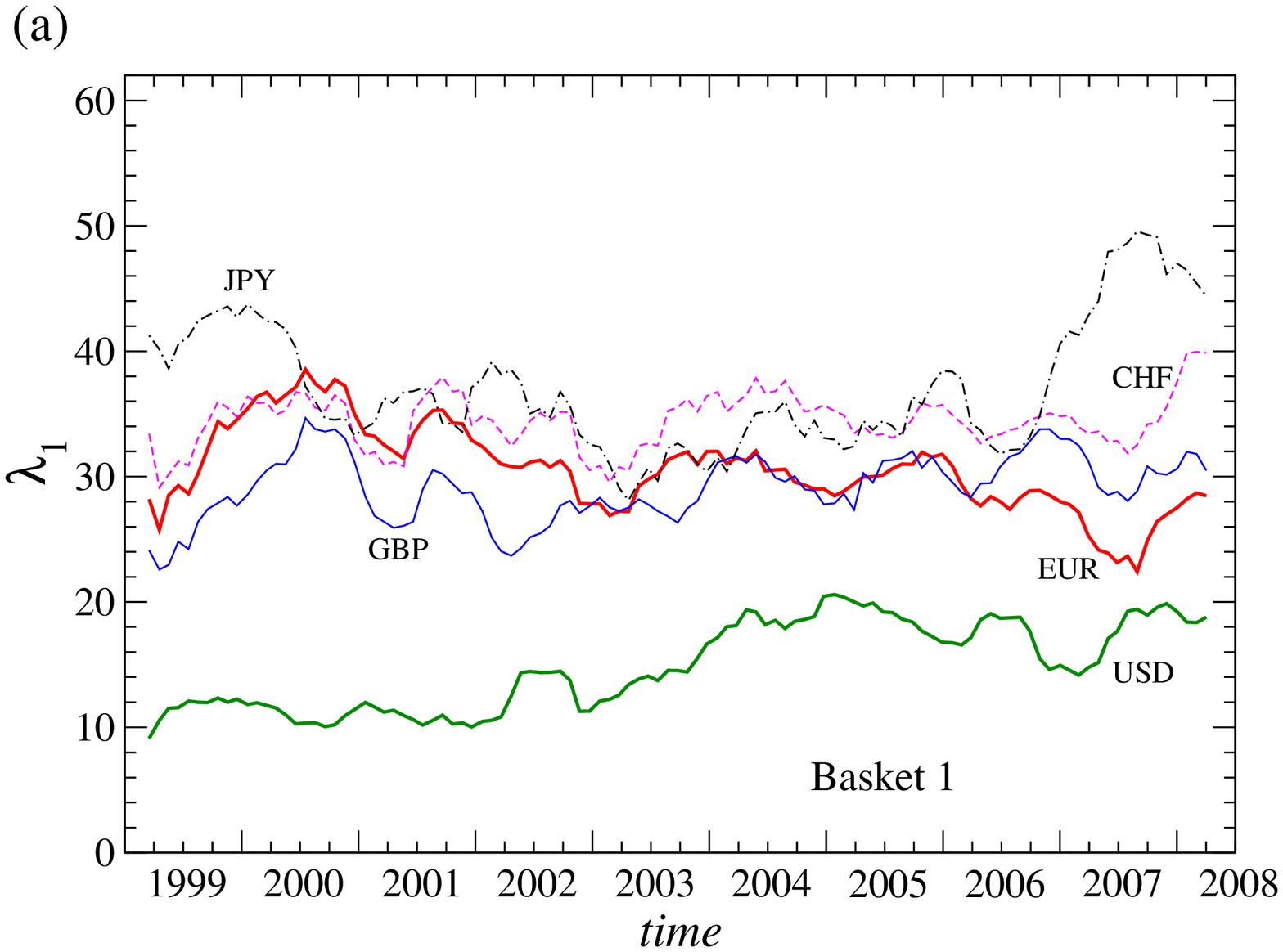}
\hspace{-1.0cm}
\epsfxsize 7cm
\epsffile{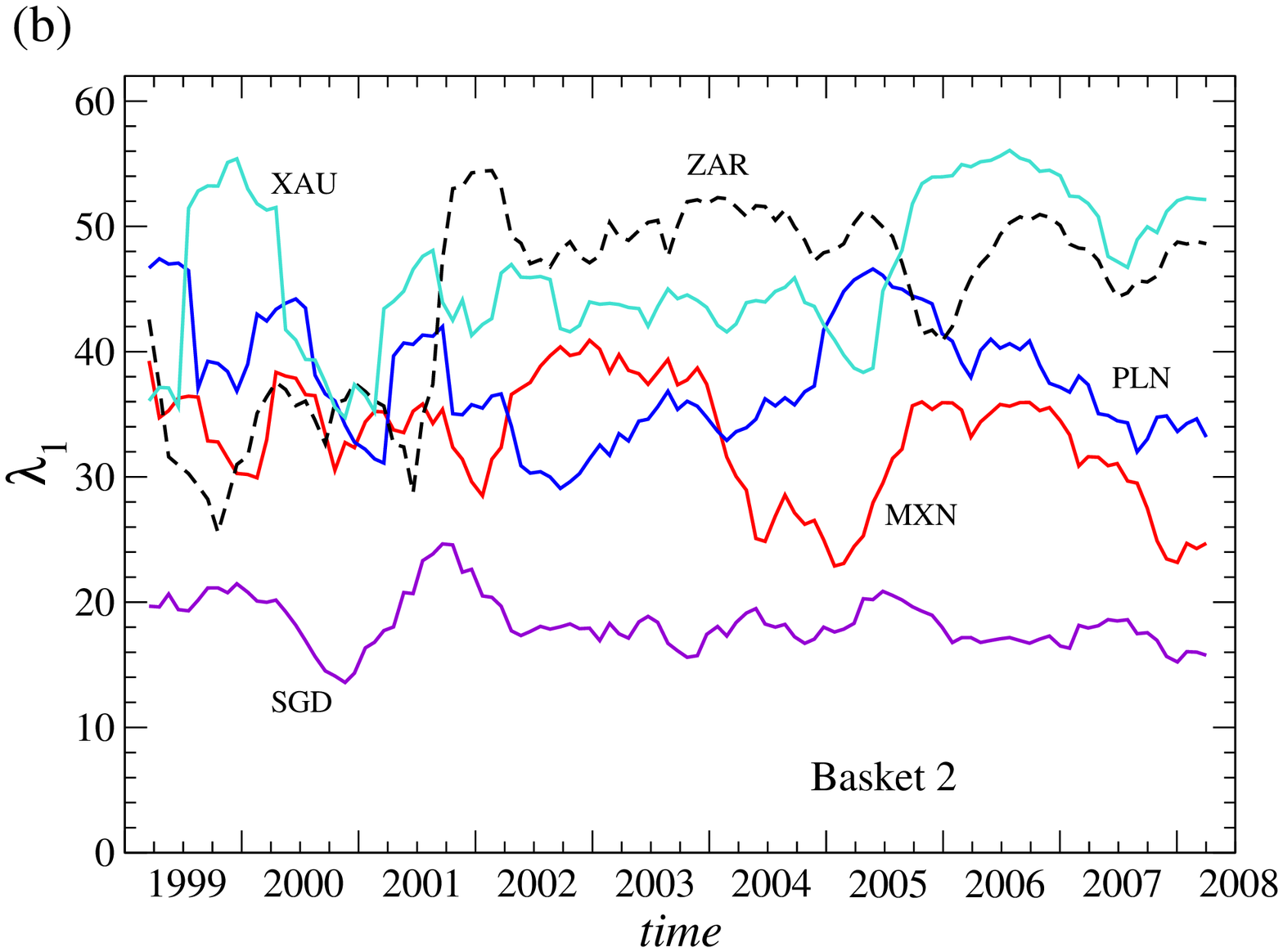}

\vspace{-0.3cm}
\hspace{2.5cm}
\epsfxsize 7cm
\epsffile{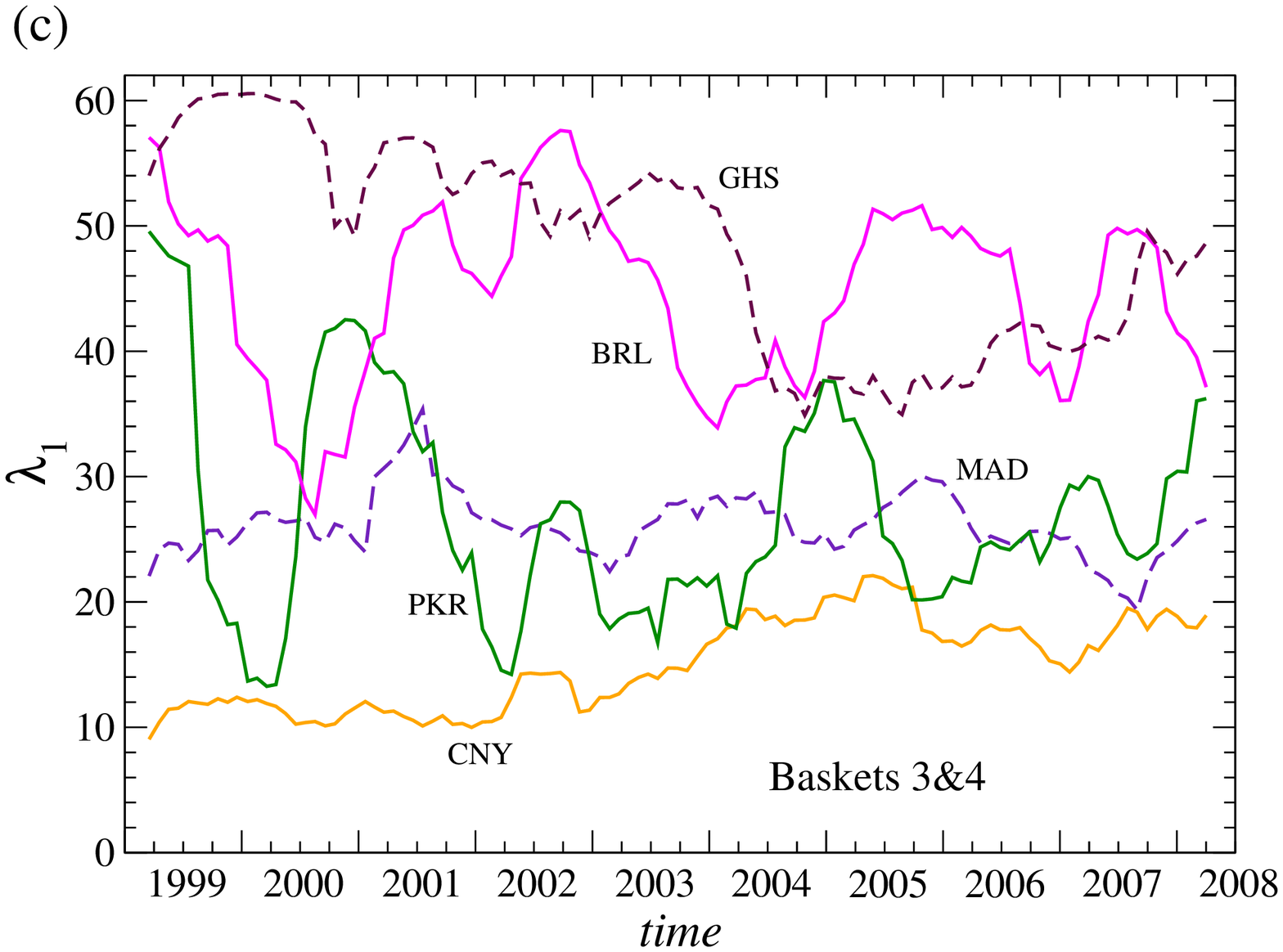}
\caption{Time dependence of $\lambda_1^{\rm B}$ for exemplary major
currencies from Basket 1 (a), other liquid currencies from Basket 2 (b),
and illiquid currencies representing Baskets 3 and 4 (c). Eigenspectra of
correlation matrices were calculated for each position of a moving window
of length of 6 months.}
\end{figure}

Figure 9 shows cumulative distributions of the node degrees for two
different network representations and for the same two intervals of time.  
A difference of the MST structures can be seen for both Bs. For Interval 1
the cumulative distribution has fatter tails while for Interval 2 the
c.d.f. decreases more quickly and develops thinner tails.  It is debatable
whether we observe any scaling behaviour of the tails, but, nevertheless,
in order to describe the different shapes of the distributions we
calculate the optimal scaling exponents. For Interval 1 we obtain
$\alpha^{\rm GBP} = 1.50$ and $\alpha^{\rm EUR} = 1.42$, while the
analogous numbers for Interval 2 are equal to 1.54 and 1.59, respectively.
In both representations there is a clear increase of the scaling 
exponent's value.

Quantitative characterization of the changes in the MST network structure 
with time can also be possible by observing the temporal variability of 
the largest eigenvalue $\lambda_1^{\rm B}$ for different choices of B. We 
divide our 9.5-years-long time interval into shorter annual periods of 
approximately 250 trading days (with an exception for the last period of 
2008, which is only 6 months long) and calculate the correlation matrices 
${\bf C}^{\rm B}$ and their eigenspectra for each period and for each base 
currency. Results are shown in Figure 10. While the upper edge of the 
eigenvalue ladders remains almost unchanged over time with $\lambda_1^{\rm 
B}$ for one or more base currencies approaching 60, their lower ends 
reveal a systematic tendency to increase from $\lambda_1^{\rm min} \simeq 
10$ in 1999 to $\lambda_1^{\rm min} \simeq 16$ in 2008. This means that 
from the perspective of USD, occupying one of the lowermost rungs of the 
ladder, the global market is nowadays more collective than it used to be 
7-9 years ago. This, in turn, means that the set of currencies which 
previously were tightly related to the US dollar, now seems to be less 
populated. Interestingly, although USD permanently remains at the bottom 
of the ladder, sometimes it loses its extreme position to the advantage of 
other currencies like HKD or SGD. However, even then USD remains the 
currency with the lowest $\lambda_1$ among all the major ones from Basket~1.

Since during the analyzed period both $\lambda_1^{\rm USD}$ and 
$\lambda_1^{\rm EUR}$ significantly changed their values indicating a lack 
of stability of the network in different representations (Figure 10), it 
is worthwhile to inspect the behaviour of the largest eigenvalue for these 
and other base currencies with a better temporal resolution. We improved 
it by applying a moving window of length of about 6 months (126 trading 
days) which was shifted by 1 month (21 trading days).

The corresponding behaviour of $\lambda_1^{\rm B}(t)$ for 15 exemplary 
currencies representing different baskets is presented in three panels of 
Figure 11. The most interesting observation is that a distance 
$\lambda_1^{\rm EUR}(t) - \lambda_1^{\rm USD}(t)$ gradually decreases 
(Figure 11(a)). This effect is caused predominantly by a systematically 
decreasing value of $\lambda_1^{\rm EUR}$, which from a magnitude of 38 in 
2000 reached a level of 23 at the end of 2007. At the same time 
$\lambda_1^{\rm USD}$, after a significant increase between 2002 and 2004, 
presently oscillates without any systematic trend between 14 and 20. On 
the other hand, the two other European major currencies: CHF and GBP, 
while sometimes closely mimicking the transient short-term behaviour of 
$\lambda_1^{\rm EUR}(t)$, do not follow its long term evolution. JPY is a 
rather different case: after a decreasing trend of $\lambda_1^{\rm JPY}$ 
between 1999 and 2003 and after a period of stabilization in 2004-2006, 
JPY now displays strong oscillations which elevated its largest eigenvalue 
to a level typical for less influential currencies (Baskets 2-4) that are 
decoupled from the global market. Figures 11(b)-(c) show $\lambda_1^{\rm 
B}$ for a few liquid and illiquid currencies from Baskets 2, 3 and 4. A 
characteristic property of this group of currencies is strong short-term 
variability of the largest eigenvalue which can be seen for almost all 
choices of B except SGD and CNY.

\section{Summary}

We presented outcomes of a study of the FX network structure based on
daily data collected for the interval 01/1999-06/2008. These outcomes 
allow us to draw the following principal conclusions:

(i) The currency network structure depends on a choice of base currency 
and the associated reference frame. On one hand, a network based on a 
currency which is decoupled from the rest of the currencies and display an 
independent behaviour shows a highly correlated, rigid structure. On the 
other hand, a network viewed from the USD perspective (or the perspectives 
of its satellites) has a richer structure with less correlations and more 
noise. For typical currencies the networks has intermediate structure that 
can be classified between these two extremes. However, for a vast majority 
of currencies, the MST graphs share the same topology quantified in terms 
of the node degree distribution. We found that these networks show a 
signature of scale free networks. An extreme opposite case are the 
USD-based, CNY-based and HKD-based networks which have topology deviating 
from scale-free in direction of a random network.

(ii) From a perspective of a typical currency, the FX network is dominated
by two strong clusters of nodes related to USD and to EUR. The former
comprises usually the Latin American and the South-East Asian currencies,
while the latter consists of the European and the Maghreb ones. There are
also other smaller groups of nodes forming clusters related to
geographical or economical factors, but normally they are masked by the
dominating two clusters and can be seen in full detail only after removing
the USD and EUR nodes from the network. Among those secondary clusters we
distinguish the cluster of Middle East currencies, the cluster of
Canadian, Australian and New Zealand dollars (sometimes accompanied by the
South African rand, which couples currencies involved in trade of various
commodities), the cluster of Central European currencies, and the cluster
of precious metals. Weaker links couple also the Scandinavian currencies,
the Latin American currencies, and the South-East Asian currencies.

(iii) We found that the FX network is not stable in time. Over a few past
years the currency network underwent a significant change of its structure
with the main activity observed in the neighbourhood of USD and EUR. The
USD-related cluster released its ties, allowing some nodes to acquire more
independence. At the same time the USD-based network becomes more
correlated, what is a different manifestation of the same phenomenon. On
the other hand, the EUR node now attracts more nodes than before and,
complimentarily, the EUR-based network reveals decreasing strength of
couplings. This might be a quantitative evidence that after a transient
period in which the FX market actors treated the new European currency
with a little of suspense, now more and more of them start to rely upon
it. These findings open an interesting topic for future research.

\end{document}